\ifdefined\pdfoutput\pdfoutput=1\fi
\PassOptionsToPackage{unicode}{hyperref}
\PassOptionsToPackage{hyphens}{url}
\PassOptionsToPackage{dvipsnames,svgnames,x11names}{xcolor}
\documentclass[
  11pt,
]{article}
\usepackage{amsmath,amssymb}
\usepackage{iftex}
\ifPDFTeX
  \usepackage[T1]{fontenc}
  \usepackage[utf8]{inputenc}
  \usepackage{textcomp} 
\else 
  \usepackage{unicode-math} 
  \defaultfontfeatures{Scale=MatchLowercase}
  \defaultfontfeatures[\rmfamily]{Ligatures=TeX,Scale=1}
\fi
\usepackage{lmodern}
\ifPDFTeX\else
\fi
\IfFileExists{upquote.sty}{\usepackage{upquote}}{}
\IfFileExists{microtype.sty}{
  \usepackage[]{microtype}
  \UseMicrotypeSet[protrusion]{basicmath} 
}{}
\makeatletter
\@ifundefined{KOMAClassName}{
  \IfFileExists{parskip.sty}{%
    \usepackage{parskip}
  }{
    \setlength{\parindent}{0pt}
    \setlength{\parskip}{6pt plus 2pt minus 1pt}}
}{
  \KOMAoptions{parskip=half}}
\makeatother
\usepackage{xcolor}
\usepackage[margin=1in]{geometry}
\usepackage{longtable,booktabs,array}
\usepackage{calc} 
\usepackage{etoolbox}
\makeatletter
\patchcmd\longtable{\par}{\if@noskipsec\mbox{}\fi\par}{}{}
\makeatother
\IfFileExists{footnotehyper.sty}{\usepackage{footnotehyper}}{\usepackage{footnote}}
\makesavenoteenv{longtable}
\usepackage{graphicx}
\makeatletter
\def\maxwidth{\ifdim\Gin@nat@width>\linewidth\linewidth\else\Gin@nat@width\fi}
\def\maxheight{\ifdim\Gin@nat@height>\textheight\textheight\else\Gin@nat@height\fi}
\makeatother
\setkeys{Gin}{width=\maxwidth,height=\maxheight,keepaspectratio}
\makeatletter
\def\fps@figure{htbp}
\makeatother
\providecommand{\tightlist}{%
  \setlength{\itemsep}{0pt}\setlength{\parskip}{0pt}}
\newlength{\cslhangindent}
\newlength{\csllabelwidth}
\newlength{\cslentryspacingunit} 
\newenvironment{CSLReferences}[2] 
 {
  \setlength{\parindent}{0pt}
  \ifodd #1
  \let\oldpar\par
  \def\par{\hangindent=\cslhangindent\oldpar}
  \fi
  \setlength{\parskip}{#2\cslentryspacingunit}
 }%
 {}
\usepackage{calc}

\usepackage{amssymb}
\usepackage{booktabs}
\ifLuaTeX
  \usepackage{selnolig}  
\fi
\IfFileExists{bookmark.sty}{\usepackage{bookmark}}{\usepackage{hyperref}}
\IfFileExists{xurl.sty}{\usepackage{xurl}}{} 
\hypersetup{
  pdftitle={The Last Costly Signal: How Generative AI Collapses Competence Signaling and Why Liability Sustains Markets for Expert Services},
  colorlinks=true,
  linkcolor={blue},
  filecolor={Maroon},
  citecolor={Blue},
  urlcolor={Blue},
  pdfcreator={LaTeX via pandoc}}

\title{The Last Costly Signal: How Generative AI Collapses Competence
Signaling and Why Liability Sustains Markets for Expert Services}
\author{Andreas Bauer\footnote{\textbf{Author note.} Aegis Compliance
  and Strategies OÜ, Tallinn, Estonia. Correspondence:
  a.bauer@science.us.org. ORCID: 0000-0002-6539-9353. \emph{Declarations
  of interest:} The author is Managing Director and Founder of Aegis
  Compliance and Strategies OÜ, a compliance and strategy advisory firm,
  and is preparing a practitioner book that draws on the framework
  developed here; both are disclosed in the interest of transparency.
  Simulation code and all numerical results are openly available at
  https://doi.org/10.6084/m9.figshare.33106946. I thank an anonymous
  reader for comments that prompted the institutional extension in
  Section 4, further rounds of critical comments for prompting the power
  analysis of Appendix C, the coverage-boundary discussion in Section
  4.2, and the robustness discussions in Sections 3.1 and 6.2, and an
  adversarial audit of the formal claims for the corrections and scope
  restrictions in this revision.}\\
Aegis Compliance and Strategies OÜ}
\date{Working Paper --- this version: August 25, 2026}

\begin{document}
\maketitle
\begin{abstract}
Generative artificial intelligence has reduced the cost of producing
convincing artifacts of expertise---reports, analyses, proposals,
references to a polished body of work---to nearly zero. Signaling theory
makes a sharp prediction for this case: signals whose informational
content rests on their production cost lose that content when production
becomes cheap. We formalize this prediction for markets for expert
services, a class of credence goods, by modeling generative AI as a
compression of the \emph{discernible headroom} between what machines
produce at negligible cost and what buyers can distinguish at all. Below
a critical headroom, no separating equilibrium in production-side
signals exists, and the information artifacts can still carry shrinks
with the headroom to zero; in the resulting pooling, high-competence
providers earn no premium, and those with outside options
exit---Akerlof's lemons dynamic. We then show that an outcome-contingent
signal---a warranty backed by damages \(D\) with ex-post verifiability
\(\varphi\)---restores full separation at \textbf{any} level of AI
capability whenever \(\varphi D \geq v\), where \(v\) is the value of a
solved problem, under four institutional preconditions stated explicitly
and then priced one by one in the paper's institutional sections: the
seller holds no private information beyond his type; the collectible
retention behind the promise is verifiable; the buyer influences neither
outcome nor claim; and enforcement deadweight is negligible relative to
the stakes. The expected cost of liability depends on whether the
problem is actually solved, not on how cheaply documents are produced;
it is therefore invariant to AI capability and preserves single crossing
when all production-side signals fail. A further proposition shows that
provenance certification priced as a type-independent stamp on unchanged
production (e.g., C2PA content credentials) cannot restore full
separation---informative mixed equilibria exist but are sender-dominated
under a maintained selection convention---while a verified commitment to
forgo the AI frontier is a different instrument: it re-imposes the
pre-AI artifact cost function and restores separation at the old
real-resource burn. Monte-Carlo simulations of an agent-based market
illustrate the dynamics: after an AI capability shock, the high type's
premium collapses (in our baseline from \(0.20\,v\) to zero) and
high-type participation falls from 88\% to 68\% when only
production-side signals are available, while both are fully sustained
when liability contracts are offered. Two further results endogenize the
institutions behind the contract: because civil-procedure costs contain
a fixed block while recoveries scale with the claim, liability signaling
has a \emph{minimum ticket size} \(v_{\min}\) below which the modeled
court-enforced warranty cannot sustain separation and the market for
small expert engagements pools; and with liability insurance, the
signal-effective quantity is not the damages promised but the risk the
seller retains or is individually priced on---under pooled premiums a
warranty separates only if the \emph{deductible} satisfies the original
condition, while experience rating restores it slowly, on the insurer's
information timescale. We state the falsification conditions of the
theory and propose a preregistered choice-based conjoint experiment with
business decision-makers in the German-speaking B2B expert-services
market---the empirical intersection (liability signal × AI disclosure ×
B2B expertise × willingness to pay) that, to our knowledge, no existing
study covers.
\end{abstract}

\textbf{Keywords:} signaling theory; credence goods; generative
artificial intelligence; service guarantees; liability; professional
services; adverse selection

\textbf{JEL codes:} D82, D86, L15, L84, M31, O33

\emph{Status of this document: working paper; this version: 25 August
2026. The theoretical propositions are proven in Appendix A; the
simulation study is illustrative of the model and is not an empirical
test. The empirical test is specified in Section 7 and has not yet been
conducted.}

\hypertarget{introduction}{%
\section{Introduction}\label{introduction}}

Two findings about the market consequences of generative artificial
intelligence appear, at first sight, to point in opposite directions.

The first is a price and quantity decline. Hui, Reshef, and Zhou (2024)
document, in a peer-reviewed difference-in-differences study of a large
online freelancing platform after the release of ChatGPT, a 2\% drop in
the number of contracts and a 5.2\% drop in monthly earnings in affected
occupations---with the additional finding that the \emph{top-rated}
providers were hit at least as hard as the rest. Non-peer-reviewed
evidence points the same way for entry-level knowledge work:
Brynjolfsson, Chandar, and Chen (2025, working paper) report an
employment decline of roughly 16\% relative to trend for
22-to-25-year-olds in the most AI-exposed occupations. In the
German-speaking consulting market, the industry association BDU (2025)
reports calculated daily rates declining for the first time in years, by
around 2\%.

The second finding is a \emph{trust discount} on machine involvement.
Schilke and Reimann (2025) show across thirteen preregistered
experiments that disclosing AI involvement in the production of work
reduces trust in the producer---robustly across professional roles, and
even when disclosure is mandatory; the effect operates through
diminished perceptions of legitimacy and goes beyond familiar algorithm
aversion. Buder and Unfried (2024) find, in a study surveying 1,000
respondents in each of the USA, the UK, and Germany (with the labeling
experiment itself fielded in Germany, \(n = 298\)), that an identical
advertisement labeled as AI-generated is rated less
favorably---particularly on emotional dimensions---than the same
material presented as a photograph. Rix, Berger, Hess, and Rzepka (2025)
document an ``algorithm discount'' in consumers' valuation of digital
products; Sikhondze, Ye, Wang, Zhan, and Santhanam (2025) find that
disclosing generative-AI production lowers willingness to pay through a
loss of perceived psychological value.

Falling prices for expert output; rising discounts on machine
involvement---declining value of what became abundant and a premium on
what remained scarce. We argue that these are not two trends but
\emph{two sides of a single comparative static}, and that the underlying
mechanism has been standard economics for half a century: \textbf{a
signal carries information only if it is costly to fake} (Spence, 1973;
Connelly, Certo, Ireland, \& Reutzel, 2011; Connelly, Certo, Reutzel,
DesJardine, \& Zhou, 2025). Generative AI has collapsed the cost of
producing the \emph{artifacts} by which expertise has traditionally been
signaled---the fluent report, the polished proposal, the comprehensive
analysis. Signals whose entire informational content rested on their
production cost are losing that content. And precisely because they are
losing it, the \emph{relative} informational value of signals that
remain expensive to fake is rising.

This paper makes four contributions.

\textbf{First, a classification.} We partition competence signals by the
\emph{locus} of their cost: \textbf{production-side signals}, whose cost
is incurred in producing an artifact before the transaction (documents,
credentials presentation, demonstrations of polish), and
\textbf{outcome-contingent signals}, whose cost is a contingent claim on
the transaction's outcome (warranties, damages, penalty
clauses---liability in the broad sense). The singular of our title
denotes this \emph{class}, not a single instrument: contingent fees and
genuinely retained insurance risk (Section 4.2) belong to it alongside
the warranty, which we analyze as its cleanest representative.
Generative AI is, in this classification, a technology that deflates the
cost of the first class and leaves the second class untouched: no
language model can reduce the expected cost of a promise to pay if the
client's problem remains unsolved. That cost depends on the provider's
actual competence, which is exactly what makes it informative.

\textbf{Second, a formal result.} We model the market for expert
services as a signaling game over a credence good (Darby \& Karni, 1973;
Dulleck \& Kerschbamer, 2006) and model generative AI as compressing the
\emph{discernible headroom} \(\eta\)---the gap between the artifact
quality achievable at negligible cost and the maximum quality buyers can
distinguish at all. Proposition 1 shows that below a critical headroom
\(\eta^{*}\), no separating equilibrium in production-side signals
exists, for essentially the reason Spence identified: the low type's
cost of mimicry falls below the price premium that separation would
command---and its part (ii) caps what any partially informative
equilibrium can still transmit. Proposition 2 traces the consequence in
the pooling benchmark that part (ii) makes exact: pooled prices, and
exit of high-competence providers with outside options---the
adverse-selection dynamic of Akerlof (1970). Proposition 3 is the core
result: a warranty backed by damages \(D\), verifiable ex post with
probability \(\varphi\), restores full separation for \emph{any} level
of AI capability whenever \(\varphi D \geq v\), where \(v\) is the value
of the solved problem. The condition has a plain reading: the promise
must cover the client's value at risk, scaled up by the imperfection of
ex-post verification. Proposition 4 shows that the same shock that
destroys production-side signaling \emph{raises} the equilibrium return
to the liability signal, because AI-enabled entry worsens the buyer's
prior---the ``scissors'' is one movement, not two. Proposition 5
establishes that signals whose cost is type-independent---including
provenance certification of human origin, such as C2PA content
credentials---cannot restore separation, no matter how cheap and
reliable certification becomes: provenance certifies \emph{process}, and
process is not competence. Propositions 6 and 7 then endogenize the
institutions behind the contract technology: courts and insurers.
Liability signaling turns out to have a minimum ticket size, set by the
fixed-cost block of civil procedure (Proposition 6), and under liability
insurance the signal-effective quantity is the risk the seller retains
or is individually priced on---not the damages promised (Proposition 7).

\textbf{Third, a simulation study.} We implement the model as an
agent-based Monte-Carlo simulation with finite populations, adaptive
participation, and low-type experimentation, and show that the
analytical results survive finite-sample noise and learning dynamics. In
the baseline calibration---parameters are stylized, and we are explicit
that the simulation \emph{illustrates} the model rather than testing
it---the AI shock erases a 20-percentage-point competence premium and
pushes roughly a fifth of high-competence providers out of the market
when only production-side signals are available; when liability
contracts are available, the premium and participation are fully
sustained, and mimicry by low types is self-extinguishing because it is
priced by realized outcomes.

\textbf{Fourth, an empirical agenda with falsification conditions.} The
mechanism is classical and we claim no novelty for it. What is genuinely
open, to the best of our knowledge after systematic search, is the
\emph{empirical intersection} of four elements: (i) liability as a
costly signal, (ii) contrast with disclosed AI production, (iii)
professional B2B expert services, and (iv) monetary willingness to pay.
Each pairwise combination has literature; the four-way intersection does
not. Section 7 specifies a preregistered choice-based conjoint
experiment with 450 business decision-makers in the German-speaking
market---a sample size derived by simulation in Appendix C rather than
by convention---that would fill it, and states what would falsify the
theory: if expert services offered with explicit warranty and liability
assumption command no price premium over several years while prices of
pure information services hold, the model is wrong.

Three honesty notes before proceeding, because a paper about trust
cannot afford less. (1) The evidence for the ``falling curve'' is not
unanimous: Humlum and Vestergaard (2025, working paper) find precisely
estimated null effects of AI chatbot adoption on earnings and hours
among \textasciitilde25,000 Danish workers in exposed occupations. Our
model is consistent with such heterogeneity---the collapse result
applies only where the service's value proposition rests on artifacts
whose quality buyers cannot verify directly---but the boundary is an
assumption of scope, not a prediction we have tested. (2) The evidence
on liability's weight in purchasing decisions is not unanimous either:
von Wedel and Hagist (2022) find in a discrete-choice experiment that
liability coverage was \emph{not} a decisive attribute for German
radiologists purchasing AI tools. Different market, different good---but
a warning against assuming rather than measuring the premium. (3) All
willingness-to-pay evidence we cite, and the experiment we propose,
measure \emph{stated} preferences; stated premiums typically overstate
revealed ones, a caveat we carry through Section 7.

The paper proceeds as follows. Section 2 locates the argument in the
relevant literatures and states the research gap precisely. Section 3
develops the baseline model and results. Section 4 endogenizes the
institutional foundations---courts and insurers---of the liability
signal. Section 5 reports the simulation study, including the
institutional modules. Section 6 discusses implications and boundary
conditions. Section 7 specifies the proposed empirical program---the
preregistered experiment and two observational implications---and the
falsification conditions. Section 8 concludes. Proofs are in Appendix A;
simulation details in Appendix B.

\hypertarget{related-literature-and-the-research-gap}{%
\section{Related Literature and the Research
Gap}\label{related-literature-and-the-research-gap}}

\hypertarget{signaling-and-its-cost-foundation}{%
\subsection{Signaling and its cost
foundation}\label{signaling-and-its-cost-foundation}}

The economics of signaling begins with Spence (1973): in markets with
asymmetric information, an observable action can transmit information
about an unobservable attribute if and only if its cost differs across
types in the right way---informativeness \emph{is} differential cost.
The management literature has absorbed and extended this insight; the
reviews by Connelly et al.~(2011) and Connelly et al.~(2025) document
its status as one of the most widely applied theoretical frames in the
field, and formulate the property we build on: signal effectiveness
resides in signal cost, and the best signals are those that are
disproportionately costly for low-quality actors. Kirmani and Rao (2000)
organize the marketing applications, distinguishing default-independent
signals (costs incurred regardless of outcome, e.g., advertising
expenditure) from \textbf{default-contingent signals} (costs incurred
only upon failure, e.g., warranties); Boulding and Kirmani (1993)
provide early experimental evidence that warranties are read as quality
signals when the seller's credibility makes the promise binding. Our
production-side vs.~outcome-contingent classification is the
Kirmani--Rao distinction pushed to its limiting case by a technology
that drives one class's cost to zero: to our knowledge the
\emph{comparative static in AI capability}---which signals survive as
artifact costs vanish---has not been formally worked out, and that, not
the mechanism, is what this paper adds.

\hypertarget{credence-goods-and-the-liability-result}{%
\subsection{Credence goods and the liability
result}\label{credence-goods-and-the-liability-result}}

Expert services are the paradigmatic credence good: the buyer cannot
fully judge quality even after purchase (Darby \& Karni, 1973). Dulleck
and Kerschbamer (2006) synthesize the theory of such markets---doctors,
mechanics, experts, consultants---and identify the institutional
conditions under which they function. The result most important for us
is experimental: Dulleck, Kerschbamer, and Sutter (2011) show in the
laboratory that \textbf{liability}---the seller being bound to actually
solve the buyer's problem---is the institution that most reliably
disciplines expert markets, more effective than increased competition
and more effective than mere verifiability of the bill. Our Proposition
3 can be read as a signaling-theoretic complement to their moral-hazard
result---with one sign flip made explicit in Section 3.3: their
liability obliges the expert to solve the problem, aligning buyer and
institution, whereas a pure damages transfer at \(\varphi D \geq v\)
sets the buyer's marginal value of success weakly negative, which is
what the split-award instruments of Section 4 repair: liability not only
disciplines behavior \emph{within} a transaction; under AI-compressed
artifact costs it becomes the \emph{only} surviving carrier of
information about competence \emph{across} transactions. Akerlof (1970)
supplies the dynamic that makes this urgent---when buyers cannot
distinguish quality, prices converge to the average and good providers
exit---and, notably, his list of ``counteracting institutions''
(guarantees, brand names, licensing, reputation) reads as a list of
costly signals ranked by how AI-resistant their cost structure is.

Because Section 4 endogenizes enforcement and insurance, two further
literatures become load-bearing. The economics of legal disputes
supplies the machinery for when a damages promise is actually
collectible: suit-and-settlement models under alternative
cost-allocation rules (Shavell, 1982; the synthesis in Cooter \&
Rubinfeld, 1989) imply that a claim is an asset only when the expected
judgment clears the cost of pursuing it, and Shavell (1986) shows that a
defendant's limited wealth truncates liability's incentive effects---the
judgment-proof problem. The economics of insurance under adverse
selection (Rothschild \& Stiglitz, 1976) supplies the screening
logic---menus of premiums and deductibles that sort risks---which
Section 4.2 turns around: the insurer becomes a delegated assessor of
expert competence, and the contract form of the insurance determines how
much of the liability signal survives.

\hypertarget{the-empirical-record-on-ai-and-expert-market-prices}{%
\subsection{The empirical record on AI and expert-market
prices}\label{the-empirical-record-on-ai-and-expert-market-prices}}

On the falling curve, the peer-reviewed anchor is Hui et al.~(2024),
discussed above; the finding that top-rated freelancers lost
disproportionately is exactly what a signal-collapse account predicts
(the premium compresses toward the pool) and harder to reconcile with a
pure productivity account. Brynjolfsson et al.~(2025, working paper,
version of November 2025) provide payroll-based evidence of entry-level
displacement in exposed occupations; we cite it as suggestive, not
settled. Against both stands the null result of Humlum and Vestergaard
(2025, working paper): AI chatbot adoption in Denmark produced precisely
estimated zero effects on earnings and hours in 2023--2024, with
self-reported time savings of about 2.8\%. We treat this not as an
anomaly to explain away but as a scope condition to respect: the
signal-collapse mechanism operates where output artifacts are the value
proposition, not wherever ``knowledge work'' happens. Labour-market
macrodata through 2026 sharpen rather than blur this reading: PwC's
\emph{2026 AI Jobs Barometer}---an analysis of over one billion job
advertisements across 27 countries; an industry study, not
peer-reviewed---reports a 62\% wage premium for workers with AI skills
and a ``two-track'' market in which roles that AI
\emph{professionalizes} grow twice as fast as roles it
\emph{democratizes}, with 42\% faster wage growth since 2021, while
AI-exposed entry-level postings increasingly demand traditionally senior
skills such as judgment and leadership (PwC, 2026). Rising pay for
identified, judgment-bearing roles and falling prices for artifact-based
output are not competing facts; in the model they are the two blades of
Proposition 4's scissors---and neither series measures the price of
credence-good artifacts directly, which is why Section 7's experiment,
not the macrodata, carries the identification.

\hypertarget{ai-disclosure-trust-and-willingness-to-pay}{%
\subsection{AI disclosure, trust, and willingness to
pay}\label{ai-disclosure-trust-and-willingness-to-pay}}

On the rising curve, Schilke and Reimann (2025) is the strongest
evidence: thirteen experiments, a robust trust penalty for disclosed AI
involvement, mediated by perceived legitimacy, surviving
mandatory-disclosure framings. Consumer-side studies concur: Buder and
Unfried (2024) on labeled advertising; Rix et al.~(2025) on the
algorithm discount in digital products; Sikhondze et al.~(2025) on
willingness to pay for disclosed GenAI production. A small preprint
reports that a majority of surveyed students state a willingness to pay
more for demonstrably human creative work (Sultana, Islam, Hossain, \&
Wasi, 2026, preprint; n = 70, convenience sample)---we cite it only as
an observation from adjacent markets, not as support. Crucially, all of
this evidence concerns consumer or creative contexts, and the signal
studied is \emph{origin} (human vs.~machine), not \emph{liability}.
McColl, Truong, and La Rocca (2019) provide the closest B2B evidence:
conjoint studies in the testing, inspection, and certification industry
showing that service guarantees carry a price premium and can ground
positioning in B2B markets---but without any AI contrast.

\hypertarget{the-gap-stated-precisely}{%
\subsection{The gap, stated precisely}\label{the-gap-stated-precisely}}

Table 1 organizes the intersection. Each element of the four-tuple---(i)
liability as the signal, (ii) AI-disclosure contrast, (iii) professional
B2B expert services, (iv) monetary willingness to pay---is covered
pairwise somewhere in the literature. We have found no study that covers
all four simultaneously. This claim rests on extensive but necessarily
non-exhaustive search and is falsifiable by a single working paper; we
state it as the current state of our knowledge, dated August 2026, not
as a certainty. The nearest recent additions---LLM sellers in
credence-goods settings (Erlei, 2025; Erlei \& Meub, 2026)---test
liability as an institutional \emph{rule} imposed on the market, not as
a signal the seller chooses, and measure no willingness-to-pay premium
for it; the four-way intersection remains open.

\begin{longtable}[]{@{}
  >{\raggedright\arraybackslash}p{(\columnwidth - 8\tabcolsep) * \real{0.4194}}
  >{\centering\arraybackslash}p{(\columnwidth - 8\tabcolsep) * \real{0.1290}}
  >{\centering\arraybackslash}p{(\columnwidth - 8\tabcolsep) * \real{0.1290}}
  >{\centering\arraybackslash}p{(\columnwidth - 8\tabcolsep) * \real{0.1935}}
  >{\centering\arraybackslash}p{(\columnwidth - 8\tabcolsep) * \real{0.1290}}@{}}
\caption{Coverage of the four elements of the research gap in the
closest existing studies. ``---'' = not addressed. The proposed
experiment is specified in Section 7.}\tabularnewline
\toprule\noalign{}
\begin{minipage}[b]{\linewidth}\raggedright
Study (year)
\end{minipage} & \begin{minipage}[b]{\linewidth}\centering
Liability signal
\end{minipage} & \begin{minipage}[b]{\linewidth}\centering
AI contrast
\end{minipage} & \begin{minipage}[b]{\linewidth}\centering
B2B expert services
\end{minipage} & \begin{minipage}[b]{\linewidth}\centering
Monetary WTP
\end{minipage} \\
\midrule\noalign{}
\endfirsthead
\toprule\noalign{}
\begin{minipage}[b]{\linewidth}\raggedright
Study (year)
\end{minipage} & \begin{minipage}[b]{\linewidth}\centering
Liability signal
\end{minipage} & \begin{minipage}[b]{\linewidth}\centering
AI contrast
\end{minipage} & \begin{minipage}[b]{\linewidth}\centering
B2B expert services
\end{minipage} & \begin{minipage}[b]{\linewidth}\centering
Monetary WTP
\end{minipage} \\
\midrule\noalign{}
\endhead
\bottomrule\noalign{}
\endlastfoot
Dulleck, Kerschbamer \& Sutter (2011) & \(\checkmark\) & --- & (lab
market) & \(\checkmark\) \\
McColl, Truong \& La Rocca (2019) & \(\checkmark\) & --- &
\(\checkmark\) & \(\checkmark\) \\
Schilke \& Reimann (2025) & --- & \(\checkmark\) & (mixed roles) &
--- \\
Buder \& Unfried (2024) & --- & \(\checkmark\) & --- & --- \\
Rix et al.~(2025); Sikhondze et al.~(2025) & --- & \(\checkmark\) & ---
& \(\checkmark\) \\
von Wedel \& Hagist (2022) & (\(\checkmark\), not decisive) &
(\(\checkmark\), AI tool purchase) & (healthcare) & \(\checkmark\) \\
Erlei (2025); Erlei \& Meub (2026) & (\(\checkmark\), as rule, not
chosen signal) & \(\checkmark\) (LLM sellers) & (lab / agent markets) &
--- \\
\textbf{Proposed experiment (Section 7)} & \textbf{\(\checkmark\)} &
\textbf{\(\checkmark\)} & \textbf{\(\checkmark\)} &
\textbf{\(\checkmark\)} \\
\end{longtable}

\hypertarget{the-model}{%
\section{The Model}\label{the-model}}

\hypertarget{setup}{%
\subsection{Setup}\label{setup}}

A market for an expert service. There is a continuum (in the simulation:
a finite population) of sellers of two types \(\theta \in \{H, L\}\)
with population shares \(\mu_0\) and \(1-\mu_0\), \(0 < \mu_0 < 1\). A
seller of type \(\theta\) solves the buyer's underlying problem with
probability \(q_\theta\), where \(0 < q_L < q_H < 1\); write
\(\Delta q \equiv q_H - q_L\). Both probabilities are conditional on the
engagement at hand: the ordering \(q_L < q_H\) is a within-task
statement---on the \emph{same} problem, competence raises the
probability of solving it---not a claim about unconditional loss rates
across portfolios. Abler sellers who systematically attract harder
mandates can show higher raw failure rates without violating it; the
distinction returns as risk adjustment in Section 4.2. One assumption
deserves its own sentence because it departs from the credence-goods
literature this paper cites as its home: beyond his type, the seller
holds no private information about the engagement at hand---no private
diagnosis that would let him select which engagements to warrant.
Darby--Karni credence goods are defined by exactly that private
diagnosis; switching it off makes this a Spence-type quality-signaling
model of an expert service, and Section 6.5(6) prices what turns back on
when it is restored. A solved problem is worth \(v > 0\) to the buyer,
an unsolved one nothing. The service is a credence good: the buyer
observes neither \(\theta\) before purchase nor, in general, whether the
problem was solved because of or despite the service---except through
the verification channel introduced below. All agents are risk-neutral.
Buyers are competitive, so the price of a service equals the buyer's
expected value of it given posterior beliefs (as in the competitive-wage
convention of Spence, 1973)---a surplus-division convention flagged here
as identifying for the empirical mapping of Section 2.3: under
seller-side price competition the same market facts arise with unchanged
information, and Proposition 3 then restores information, not
necessarily rents. High-type sellers have outside options \(r\)
distributed with c.d.f. \(F\) on
\([\underline{r}, \bar{r}] \subset (0,\, q_H v)\); low types' outside
option is normalized to \(0\). Where a result additionally requires
\(\underline{r} > q_L v\), this is invoked explicitly (the
\(\lambda = 0\) clause of Proposition 2 and part (i) of Proposition 5).
A seller who does not participate sends no signal, trades nothing, and
earns the outside option; separation statements refer to active sellers.

\textbf{Production-side signals (artifacts).} Before contracting, a
seller can present an artifact---sample work, a diagnostic report, a
proposal, the visible polish of a body of work---of perceived quality
\(s \in [0, \bar{s}]\). The ceiling \(\bar{s}\) is a property of
\emph{buyers}: it is the highest artifact quality a non-expert buyer can
distinguish at all. Producing perceived quality \(s\) costs \[
c(s, \theta; a) \;=\; k_\theta \cdot \max\{0,\; s - a\}, \qquad k_L > k_H > 0,
\] where \(a \in [0, \bar{s}]\) is the \textbf{AI frontier}: the
artifact quality obtainable at negligible marginal cost from generative
tools. The type-dependence \(k_L > k_H\) captures that producing
\emph{convincing} expert artifacts beyond the machine frontier is harder
without the underlying competence. Define the \textbf{discernible
headroom} \[
\eta \;\equiv\; \bar{s} - a .
\] Generative AI progress is modeled as an increase in \(a\), i.e., a
compression of \(\eta\) toward zero. This is the formal rendering of an
empirical regularity documented in the disclosure literature: buyers
increasingly \emph{cannot tell} machine output from expert output within
the quality range they can perceive (Schilke \& Reimann, 2025; Buder \&
Unfried, 2024).

\textbf{Outcome-contingent signals (liability).} A seller may attach a
warranty \(w \in \{0, 1\}\) to the offer: a binding commitment to pay
damages \(D \geq 0\) to the buyer if the problem turns out unsolved.
Ex-post verification of failure succeeds with probability
\(\varphi \in (0, 1]\) (the \emph{verifiability} of the outcome, in the
sense of Dulleck \& Kerschbamer, 2006). The expected cost of the
warranty to a seller of type \(\theta\) is therefore \[
\kappa_\theta(D) \;=\; (1 - q_\theta)\,\varphi D ,
\] which is \textbf{decreasing in competence} and \textbf{independent of
\(a\)}: no improvement in generative tools changes the probability that
the client's problem remains unsolved, conditional on type. This
invariance---not any assumption about attitudes toward AI---is what
drives every result below. It is an invariance of \(\kappa_\theta\)
\emph{given} \(\varphi\): where Section 4.1 endogenizes \(\varphi\), AI
cuts both ways---cheaper drafting lowers the fixed access block and
raises credibility, cheaper defensive documentation lowers the
plaintiff's \(\rho\)---so \(\partial \varphi/\partial a = 0\) is an
assumption, not a result, and is stated as such.

Timing: (1) sellers observe their type and choose participation,
artifact level \(s\), and warranty \(w\); (2) buyers observe \((s, w)\)
and, where the contract stipulates them, verifiable collateral
terms---an escrowed retention, a certified and funded cover, a rated
premium (Section 4.2); solvency itself is never observed, which is why
collectible, verifiable retention rather than the printed promise
carries the signal there; beliefs form and prices are set competitively;
(3) outcomes realize; verified failures trigger damages. The solution
concept is perfect Bayesian equilibrium (PBE); where selection matters
we use the least-cost separating (Riley) outcome and note where the D1
refinement does the usual work.\footnote{With two types, the Intuitive
  Criterion of Cho and Kreps (1987) already eliminates pooling wherever
  a separating deviation exists; we name D1 (Banks \& Sobel, 1987) only
  because it is the standard workhorse. Nor is the two-type structure
  load-bearing: with a continuum of types, separating equilibria in a
  continuously variable signal exist under single crossing (Mailath,
  1987), and D1 selects the least-cost (Riley) outcome (Cho \& Sobel,
  1990; Ramey, 1996). The binary warranty of the baseline then supports
  a threshold equilibrium---warranty above a competence cutoff, none
  below---while a menu of damages levels \(D\) restores full separation,
  since the warranty cost \((1 - q)\varphi D\) satisfies single crossing
  in \((q, D)\). None of the comparative statics in AI capability depend
  on the cardinality of the type space, because \(a\) enters no
  liability payoff. A separate line of criticism targets selection
  \emph{across} equilibria rather than beliefs within one: the Stiglitz
  critique holds that off-path messages should be interpreted by
  comparing entire equilibria, which motivates the undefeated
  equilibrium of Mailath, Okuno-Fujiwara, and Postlewaite (1993) and,
  recently, the persuasiveness criterion of Zeng (2025, preprint). The
  liability result is robust here by structure, not by refinement:
  because the truthful sender's expected signaling cost is zero in the
  settlement benchmark---and stays below the pooling shortfall wherever
  \(q_H v - (1-q_H)\ell C_L(D) \geq \bar{p}(\lambda)\), which holds at
  large tickets and fails at small ones (Section 4.1)---the high type
  strictly prefers warranty separation to no-warranty pooling at every
  interior \(\lambda\) on that region, while the mimic's deviation to
  the warranty is equilibrium-dominated under \(\varphi D > v\) (weakly
  at the boundary); the separating outcome is therefore undefeated and
  survives cross-equilibrium selection. Artifact separation is not: for
  \(\lambda > 1 - k_H/k_L\) both types prefer the artifact pooling
  (cf.~Proposition 2), and cross-equilibrium criteria select it against
  the Riley outcome---production-side separation was fragile on this
  margin even before the AI shock, which sharpens the paper's contrast
  rather than weakening it.}

\hypertarget{the-collapse-of-production-side-signaling}{%
\subsection{The collapse of production-side
signaling}\label{the-collapse-of-production-side-signaling}}

\textbf{Proposition 1 (Signal collapse).} \emph{(i) A separating PBE in
artifact signals with active high types exists if and only if
\(F\!\left(q_H v - (k_H/k_L)\Delta q\, v\right) > 0\) (the Riley payoff
clears at least the lowest outside options) and} \[
\eta \;\geq\; \eta^{*} \;\equiv\; \frac{\Delta q\, v}{k_L}.
\] \emph{(ii) In any PBE of the artifact-signaling game---prices
conditioning on \(s\) alone---the essential spread of on-path posterior
beliefs (regular conditional beliefs at on-path signals) that a seller
is high-type is at most \(\min\{1,\ \eta / \eta^{*}\}\): the maximal
information artifacts can carry declines linearly with the headroom and
vanishes as \(\eta \to 0\).}

The logic is Spence's, relocated: separation requires the high type to
display an artifact level \(s_H\) expensive enough that the low type
declines to mimic, \(k_L (s_H - a) \geq \Delta q\, v\); feasibility
requires \(s_H \leq \bar{s}\). Both hold only if the headroom is at
least \(\eta^{*}\). The participation clause is not decoration: the
separating price must not only deter the mimic but also retain the
sender, and if every high type's outside option exceeds the Riley payoff
\(q_H v - (k_H/k_L)\Delta q\, v\), the separating profile exists on
paper while the market it describes is empty. The scope clause in (ii)
matters equally: once a second, costless-to-correlate instrument such as
the warranty of Section 3.3 is on the table, an artifact level can ride
along as a payoff-irrelevant label and carry that instrument's
information---part (ii) bounds what artifacts can transmit on their own
strength. The AI frontier does not need to reach the buyers' ceiling for
signaling to die---it needs only to come within \(\eta^{*}\) of it. Part
(ii) sharpens the point: below \(\eta^{*}\), artifacts do not merely
fail to separate fully---the most information any equilibrium can still
transmit through them is capped by \(\eta / \eta^{*}\) and decays
continuously to zero with the headroom. In what follows we take the
uninformative pooling outcome as the post-collapse benchmark; by part
(ii) it is the limit \(\eta \to 0\) and a bound on every equilibrium in
between. The linear cost family is expositional, not load-bearing: for
any strictly increasing \(c_\theta(\cdot)\) with \(c_\theta(0) = 0\) and
\(c_L > c_H\) pointwise, the same argument yields the threshold
\(\eta^{*} = c_L^{-1}(\Delta q\, v)\) and, for part (ii), the bound
\(c_L(\eta)/(\Delta q\, v)\)---the sharp cutoff and the continuous decay
of artifact informativeness are properties of the single-crossing
structure, not of linearity. Two objects are deliberately held fixed in
the comparative static: the buyers' ceiling \(\bar{s}\) and the low
type's slope \(k_L\). Both are contestable---AI-assisted evaluation
would raise \(\bar{s}(A)\), AI assistance \emph{above} the frontier
would lower \(k_L(A)\) and thus raise \(\eta^{*}(A)\), accelerating
collapse---and the kinked family makes \(\eta^{*}\) itself
\(a\)-independent by construction; the statement is a comparative static
in \(a(A)\) alone, with \((\bar{s}, k_L)\) held fixed and said so. All
proofs are in Appendix A.

\textbf{Proposition 2 (Pooling and exit).} \emph{For
\(\eta < \eta^{*}\), full separation is infeasible (Proposition 1), and
partially informative equilibria transmit at most \(\eta/\eta^{*}\) of
the type information (Proposition 1(ii)). We select the complete-pooling
benchmark---every limit point of such equilibria as \(\eta \to 0\) is
uninformative conditional on its limiting active composition, and among
pooling fixed points we track the largest---in which the market trades
at price \(\bar{p}(\lambda) = q_L v + \lambda\, \Delta q\, v\), where
\(\lambda\) is the equilibrium share of high types among active sellers,
determined as a fixed point of} \[
\lambda \;=\; \frac{\mu_0\, F(\bar{p}(\lambda))}{\mu_0\, F(\bar{p}(\lambda)) + (1 - \mu_0)} .
\] \emph{A fixed point exists (under \(\underline{r} > q_L v\),
\(\lambda = 0\), the lemons-only outcome, is always one; we track the
largest); at any fixed point with \(\lambda < 1 - k_H / k_L\), the high
type's payoff is strictly below its separating payoff, and high-type
participation is weakly below its separating level. If
\(F(q_L v + \mu_0\, \Delta q\, v) = 0\)---in particular, if
\(\underline{r} > q_L v + \mu_0\, \Delta q\, v\), or if equality holds
and \(F\) carries no atom at \(\underline{r}\)---the only fixed point is
\(\lambda = 0\) and the market unravels completely: only low types
trade, at price \(q_L v\).}

Proposition 2 delivers Akerlof's dynamic, in the pooling benchmark, as a
consequence of Proposition 1, and it matches the otherwise puzzling
detail in Hui et al.~(2024) that \emph{top-rated} providers lost the
most: a collapse of separation compresses the premium of the
identified-good, which is a loss concentrated precisely on those who
previously separated. Note what Proposition 2 does \emph{not} predict: a
collapse of average prices everywhere. In our baseline calibration the
pooled price falls only modestly; what collapses is the \emph{premium},
and what erodes is \emph{composition}.

\hypertarget{liability-restores-separationat-any-ai-capability}{%
\subsection{Liability restores separation---at any AI
capability}\label{liability-restores-separationat-any-ai-capability}}

\textbf{Proposition 3 (The last costly signal).} \emph{Maintain: (A1)
beyond \(\theta\), the seller holds no private information about the
engagement; (A2) the promised transfer is collectible and its retention
verifiable; (A3) the buyer influences neither the outcome nor the claim;
(A4) enforcement deadweight is negligible---the settlement benchmark of
Section 4.1. Suppose \(D \geq v / \varphi\). Then for every
\(\eta \geq 0\)---in particular for \(\eta = 0\), artifact signals fully
collapsed---the following is a PBE: high types offer the warranty and
trade at \(p_H = q_H v + (1 - q_H)\varphi D\); low types do not offer it
and trade at \(p_L = q_L v\); buyers believe a warranted offer comes
from \(H\). The high type's net payoff is \(q_H v\). The warranty choice
is strictly incentive compatible for \(\varphi D > v\) and weak---the
mimic exactly indifferent---at \(\varphi D = v\); in the artifact
dimension the PBE is weak throughout, all levels \(s \leq a\) being
costless and payoff-equivalent under beliefs that condition on \(w\)
alone. Existence, not uniqueness, is asserted. It is supported by the
cost differential} \[
\kappa_L(D) - \kappa_H(D) \;=\; \Delta q\, \varphi D \;\geq\; \Delta q\, v,
\] \emph{which is invariant to \(a\): single crossing in the liability
dimension survives any level of AI capability.}

Conditions (A1)--(A4) are not regularity padding; they are the
institutional checklist the rest of the paper prices. Section 6.5(6)
relaxes (A1): private diagnosis lets a mimic warrant selectively, and
the required damages diverge. The remarks below and Section 4.2 relax
(A2): unobservable solvency caps the warranty posterior below one for
every \(D\), and portfolio scale defeats per-engagement deterrence---at
Table 2, a mimic with collectible wealth \(2v\) escapes deterrence
already at two concurrent engagements, at \(10v\) at nineteen, so scale,
the very margin AI relaxes, attacks the signal, and per-engagement
collateral is the answer. The second remark relaxes (A3); Section 4.1
prices (A4). What survives all four relaxations is the classification;
the free lunch does not.

The condition \(\varphi D \geq v\) has a plain-language reading that
practitioners will recognize: \emph{the warranty must cover the client's
value at risk}---and it must cover it more the leakier the ex-post
verification of failure. With perfect verifiability (\(\varphi = 1\)),
damages equal to the value of the problem suffice; with
\(\varphi = 0.5\), the promise must be twice the value at risk to carry
the same information. This is the signaling-side complement to the
experimental finding of Dulleck et al.~(2011) that liability, not
verifiability alone and not competition, keeps expert markets
functioning: in our model verifiability enters only as a multiplier on
how much liability is needed.

Four remarks. First, the high type's \emph{posted} price rises by the
expected damages \((1-q_H)\varphi D\)---the warranty is priced, not
donated---but its \emph{net} payoff is exactly \(q_H v\): in the
frictionless settlement benchmark the damages are an actuarially fair
transfer, and the signal is costless \emph{to the truthful sender} in
expectation while being loss-making to the mimic whenever
\(\varphi D > v\) and exactly break-even at the boundary. That asymmetry
is the entire point---and the freeness, unlike the invariance to \(a\),
does not survive the paper's own institutions: with a positive
litigation rate \(\ell\), loser-pays drives a wedge of exactly
\(C_L(D)\) between what the seller expects to pay and the buyer to
receive, and the honest sender bears \((1-q_H)\,\ell\,C_L(D) > 0\)---at
\(v = €5{,}000\) and full litigation, \(16\%\) of \(v\), more than the
artifact signal ever cost (Section 4.1). Second, the same condition that
deters the mimic sets the buyer's marginal value of success to
\(v - \varphi D \leq 0\) (\(-0.125\,v\) at Table 2): on the entire
separating region the buyer weakly prefers the problem
unsolved---deterrence and two-sided moral hazard are one threshold seen
from opposite sides, and the client-side instruments of Sections 4 and
6.2 (acceptance criteria excluding client-caused failure; splitting the
award into a compensatory component \(D_c \leq v\) to the buyer and a
punitive component \(D_p\) to escrow or a third party, with
\(\varphi(D_c + D_p) \geq v\)) are part of the condition's
implementability, not refinements. Third, \(H\)'s payoff is \(q_H v\)
for every admissible \(D\): the equilibrium set is a continuum with a
single payoff, and any \(\varepsilon\) of drafting or capital cost
selects the knife edge \(D = v/\varphi\), where deterrence is only
weak---the simulation's \(D = 1.25\,v > 1.11\,v\) sits deliberately
inside the strict region. Fourth, risk aversion and wealth bound
feasible \(D\) for small firms---under CARA with coefficient \(\gamma\),
the honest sender's certainty-equivalent cost is approximately
\(\tfrac{1}{2}\gamma\, q_H(1-q_H)(\varphi D)^2\), quadratic in the
ticket, a testable prediction---and solvency is \emph{private}: a mass
\(\psi\) of unobservably judgment-proof mimics caps the warranty
posterior at \(\mu_0/(\mu_0 + (1-\mu_0)\psi)\)---\(0.95\) at
\(\psi = 0.05\)---for every \(D\), which is why indemnity insurance,
escrow, and verified retention are signaling infrastructure rather than
background (Section 4.2).

\textbf{Proposition 4 (The rising blade).} \emph{The gross
identification premium---the value to a high type of being identified
rather than pooled, before any signaling cost---is
\(\Pi(\lambda) = (1 - \lambda)\, \Delta q\, v\), strictly decreasing in
\(\lambda\). If AI capability \(A\) (weakly) raises the frontier
\(a(A)\) and (weakly) increases entry of low-type-equivalent supply, so
that \(\lambda(A)\) is (weakly) decreasing, then: (i) for \(A\) above
the threshold at which \(\eta(A) < \eta^{*}\), the full-separation
artifact rent is zero, and any residual artifact informativeness---and
with it any on-path artifact rent---is bounded through Proposition 1(ii)
by \(\Delta q\, v \cdot \min\{1,\, \eta(A)/\eta^{*}\}\)---the
dimensionless posterior bound times the stakes---and vanishes as
\(\eta(A) \to 0\); (ii) \(\Pi(\lambda(A))\) is (weakly) increasing in
\(A\). The same parameter movement destroys the production-side signal
and raises the return to the surviving signal.}

This is the ``scissors'' of the title of our broader project, as one
comparative static: the falling blade (artifact rents to zero) and the
rising blade (the identification premium grows as the pool sours) are
driven by the single parameter \(A\). Figure 1 plots both rents against
\(A\) in the baseline calibration. Two guards against overreading:
\(\Pi\) is gross---net of the Riley artifact cost
\((k_H/k_L)\Delta q\, v\), the artifact rent is
\((1 - \lambda - k_H/k_L)\,\Delta q\, v\), strictly smaller (at
\(\lambda = 0.4\) and \(k_H/k_L = 1/3\) in the baseline calibration:
\(0.18\) gross versus \(0.08\) net)---and the composition path
\(\lambda(A)\) is an assumption of this comparative static, not derived
from the participation fixed point; Section 5 endogenizes it
numerically. A third guard: \(\Pi\) is the gain from \emph{adopting} the
warranty against a pooled market, not the realized separating
premium---in the Proposition 3 equilibrium unwarranted offers are priced
\(q_L v\), so the realized premium is flat at \(\Delta q\, v\), exactly
what Section 5.2 reports; Figure 1's rising dashed line plots the
adoption incentive.

\begin{figure}
\centering
\includegraphics[width=0.88\textwidth,height=\textheight]{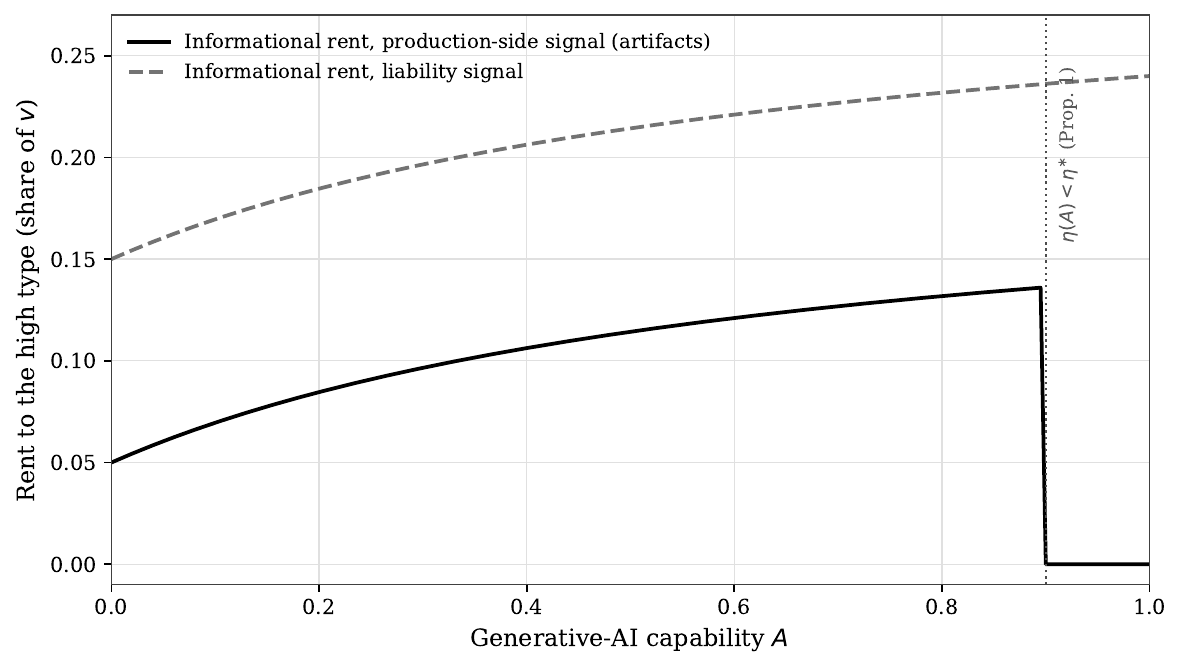}
\caption{The scissors as one comparative static. Informational rent to
the high type from production-side signals (solid) and from the
liability signal (dashed) as generative-AI capability \(A\) rises. The
production-side rent grows initially---a better pool raises the
identification premium for everyone still able to separate---then drops
to zero when the discernible headroom falls below \(\eta^{*}\)
(Proposition 1). The dashed line is the gross gain from adopting the
warranty against the pooled market (Proposition 4); the realized premium
in the separating equilibrium is flat at \(\Delta q\, v\), and the
adoption incentive is continuous and increasing throughout. Baseline
parameters of Table 2.}
\end{figure}

\hypertarget{what-certification-of-origin-cannot-do}{%
\subsection{What certification of origin cannot
do}\label{what-certification-of-origin-cannot-do}}

A prominent objection runs: verification technology will become
cheap---content credentials (C2PA specification, version 2.4) let
providers cryptographically bind provenance assertions, origin and edit
history, to their files---and Article 50 of the EU AI Act (Regulation
(EU) 2024/1689, as amended by Regulation (EU) 2026/1744) mandates
machine-readable marking of AI-generated content from 2 August
2026---generative systems already on the market before that date have
until 2 December 2026 to comply. Does cheap provenance not repair the
artifact signal?

\textbf{Proposition 5 (Provenance is not competence).} \emph{Let a
certificate cost \(m \geq 0\) identically across types and leave
\(q_\theta\) unaffected. (i) Under \(\underline{r} > q_L v\), no PBE
with a positive mass of active high types separates the types in pure
strategies through the certificate. (ii) Put \(t = m/(\Delta q\, v)\).
For every active prior \(\mu \in (0,1)\) and every aggregate
certification probability \(\omega \in (0,1]\) with
\(0 \leq \mu - \omega t\) and \(\mu + (1-\omega)t \leq 1\), the
strategies \(\alpha_H = \omega[\mu + (1-\omega)t]/\mu\) and
\(\alpha_L = \omega[1 - \mu - (1-\omega)t]/(1-\mu)\) form a
Bayes-plausible informative PBE at that prior: the certificate's price
gap equals \(m\), both types are indifferent, and the common payoff is
\(u(\mu, \omega) = q_L v + \mu\,\Delta q\, v - \omega m\). With
participation endogenous, an informative equilibrium exists exactly when
some admissible pair \((\mu, \omega)\) also solves the participation
fixed point
\(\mu = \mu_0 F(u(\mu,\omega))/(\mu_0 F(u(\mu,\omega)) + 1 - \mu_0)\)
with \(F(u(\mu,\omega)) > 0\). Existence can still fail outright:
\(u(\mu, \omega) < q_L v + \mu_0 \Delta q\, v\) strictly for every
\(\omega > 0\), so primitives with
\(\underline{r} \geq q_L v + \mu_0\, \Delta q\, v\) admit no informative
certificate equilibrium for any \(m\). (iii) Every informative
equilibrium of (ii) pays each type the uncertified price \(p(0)\),
strictly below both types' payoff in the same-population equilibrium
where nobody certifies: certification burns \(m\) to move beliefs while
making every seller strictly worse off. Every informative equilibrium of
(ii) is strictly sender-dominated, for all sellers active in it, by a
no-certificate pooling equilibrium whose active high-type share is
weakly larger; under the maintained selection convention that discards
equilibria so dominated---an assumption, not a refinement implied by the
model---posterior beliefs about competence equal the prior, conditional
on participation, in every retained equilibrium.}

The pure-strategy part is elementary---with type-independent costs, both
types face the same deviation calculus, so any belief system that
rewards the certificate is adopted by both. The mixed equilibria of part
(ii) do not rescue the certificate: they coordinate on waste, moving
beliefs at the cost of \(m\) while leaving every seller strictly worse
off than not certifying at all. The implication stands: provenance
technology certifies \emph{process} (this file was produced by a human,
this file was not altered), while the buyer's problem is
\emph{competence} (will this provider solve my problem?). A signature
can make origin cheap to verify; it cannot make competence expensive to
fake. Liability can, because its cost is written on the outcome
distribution itself. Provenance and liability are therefore not
substitutes but different objects, and policy that mandates the former
(AI Act labeling) should not be expected to restore the market
information that the latter carries. Where origin \emph{causally} raises
solution probability, origin certification inherits informativeness
through that channel---but then it is the competence embedded in the
process, not the certificate, doing the work.

One distinction the proposition must not be read past: it prices the
certificate as a \emph{stamp} on unchanged production possibilities. A
verified commitment to produce \emph{without} the frontier---``no
generative step occurred,'' which is what a content credential attests
along a declared tool chain---is a different instrument: on the
certified path the artifact does not ride free to \(a\), the pre-AI cost
function \(k_\theta \cdot s\) returns intact, and separation requires
only \(\bar{s} \geq \eta^{*}\), a condition containing no \(a\) at all.
At Table 2 this holds by a factor of ten at every frontier:
certified-human separation at \(s_H = 0.10\), high-type cost
\(0.10\,v\), payoff \(0.75\,v\), participation \(87.5\%\)---the
pre-shock equilibrium digit for digit, and a certificate fee \(m\) only
lowers the required level to \((\Delta q\, v - m)/k_L\). The honest case
against provenance is therefore not feasibility. It is threefold:
certified-human separation burns \((k_H/k_L)\Delta q\, v\) of real
resources per transaction where the settlement-benchmark warranty burns
none; it is defeated by metadata stripping and hybrid workflows; and as
AI improves, ``human-made'' and ``competent'' decouple, so the
certificate ceases to proxy anything the buyer wants. The first two are
welfare and enforcement rankings; the third is this paper's actual
thesis.

\hypertarget{institutional-microfoundations-courts-and-insurers}{%
\section{Institutional Microfoundations: Courts and
Insurers}\label{institutional-microfoundations-courts-and-insurers}}

The baseline model treats the damages \(D\) and the verifiability
\(\varphi\) as exogenous contract technology. In real markets both are
produced by institutions: \(D\) is bounded by the seller's solvency and
insurance cover, and \(\varphi\) is not a property of documents but the
outcome of a decision to litigate under a specific cost-allocation
regime. This section endogenizes both, taking the institutions of the
German-speaking market as the concrete case, and derives two results
that discipline the theory's scope: liability signaling has a
\emph{minimum ticket size} (Proposition 6), and under liability
insurance the signal-effective quantity is the risk the seller retains
or is individually priced on, not the damages promised (Proposition 7).

\hypertarget{courts-the-credible-threat-foundation-of-varphi}{%
\subsection{\texorpdfstring{Courts: the credible-threat foundation of
\(\varphi\)}{Courts: the credible-threat foundation of \textbackslash varphi}}\label{courts-the-credible-threat-foundation-of-varphi}}

A warranty is only as informative as the client's ability to collect on
it. Consider the enforcement subgame after a failure. Under the
loser-pays cost allocation of German civil procedure (§ 91 ZPO; the
Austrian rule is analogous), the client recovers damages and costs if
she prevails and bears both parties' costs if she does not. Let
\(\rho \in (0,1)\) be the probability of proving the claim. German law
helps her partway---under § 280(1) BGB fault is presumed once a breach
of duty is established, and § 1298 ABGB reverses the burden similarly in
Austria---but the breach itself, causation, and quantum remain hers to
prove, so \(\rho < 1\) even for a meritorious claim. Let the total
litigation costs of both sides be \[
C_L(D) \;=\; c_0 + c_1 D^{\alpha}, \qquad \alpha \in (0,1),\ c_0 > 0,\ c_1 > 0,
\] a stylized rendering of the degressive statutory fee schedules
(GKG/RVG): a fixed access block plus fees that grow less than
proportionally with the amount in dispute. At the stylized parameters
used below, total two-sided costs run on the order of 84\% of a €5,000
claim, 24\% of a €50,000 claim, and 10\% of a €500,000 claim---orders of
magnitude, not a fee calculation.

The threat of suit is credible if and only if the expected value of
suing is positive, \(\rho D - (1-\rho)\, C_L(D) > 0\). Under a credible
threat, rational parties settle in the shadow of the expected judgment
(Shavell, 1982; Cooter \& Rubinfeld, 1989), with expected transfer
\(\rho D\)---the risk-neutral, frictionless-bargaining benchmark; actual
settlement shares depend on protocol and bargaining power, which shifts
levels, not the threshold structure below. Absent a credible threat, the
warranty is legally a promise and economically a decoration. One further
institutional map belongs here: courts also moderate the promise itself.
German law permits judicial reduction of disproportionate contractual
penalties (§ 343 BGB), excludes that reduction between merchants (§ 348
HGB)---the core of our B2B scope---while standard-terms control applies
even there (§§ 307, 310 BGB); Austrian law mandates moderation of
excessive penalties on plea (§ 1336(2) ABGB). In the model's terms,
moderation caps the collectible amount: it lowers \(D_{\mathrm{eff}}\),
not \(\varphi\), and every condition below must hold at the moderated
level---an enforceability map \(D \mapsto D_{\mathrm{legal}}(D)\) whose
occupation- and jurisdiction-specific calibration is developed in
companion work (Bauer, 2026b). The baseline model's \(\varphi\) is
thereby micro-founded: \[
\varphi \;=\; \rho \cdot \mathbf{1}\{\text{suit credible}\}
\] ---a composite enforceability parameter (merits, fee recovery,
collection) times credibility of enforcement; if settlement transfers
\(\beta \rho D\) with \(\beta \neq 1\), deterrence reads
\(\beta\rho\delta \geq 1\) and every threshold below scales accordingly;
the fee function is calibration, not a statutory implication. The deeper
wedge is one-sided: conditional on litigation the loser pays both sides'
costs, so the seller's effective collection parameter
\(\varphi_S = \rho(1 + C_L/D)\) exceeds the buyer's
\(\varphi_B = \rho - (1-\rho)C_L/D\) by exactly \(C_L(D)/D\)---no single
\(\varphi\) prices both sides once cases are litigated rather than
settled---and the honest seller's expected cost
\((1-q_H)\,\ell\,C_L(D)\) creates a \emph{seller-side} minimum ticket of
its own: at the Table 2 composition and full litigation, the gain from
separating clears it only above roughly €4,300, four times the
buyer-side \(v_{\min}(0.60)\). We carry \(\varphi\) as the buyer-side
parameter of the settlement benchmark and flag this wedge as the
litigation extension's object. Scale damages with the ticket,
\(D = \delta v\):

\textbf{Proposition 6 (Small-ticket exclusion).} \emph{Suppose
\(\rho\,\delta > 1\). Then there exists a unique threshold
\(v_{\min}(\rho) > 0\), solving
\(\rho\,\delta\, v = (1-\rho)\, C_L(\delta v)\), such that liability
separation is feasible if and only if \(v > v_{\min}\) (at
\(v = v_{\min}\) the threat is exactly non-credible under the strict
credibility definition); \(v_{\min}\) is strictly increasing in \(c_0\)
and strictly decreasing in \(\rho\). If \(\rho\,\delta < 1\), liability
separation fails at every ticket size; at the knife edge
\(\rho\,\delta = 1\), deterrence holds with equality and (weak)
separation is feasible exactly where the threat is credible,
\(g(v) > 0\). Below \(v_{\min}\) the market is governed by Propositions
1--2, the modeled court-enforced warranty being noncredible. (Reduced
form: \(\delta > 0\), settlement at exactly \(\rho D\), and full
collectibility and legal validity of the award are assumed.)}

The deterrence part of the condition, \(\rho\delta \geq 1\), is the
familiar \(\varphi D \geq v\) with the micro-founded \(\varphi\):
because the mimic is punished only in expectation of a \emph{provable}
failure, the contractual promise must gross up for the burden of
proof---\(\delta \geq 1/\rho\), a penalty multiple, which is why
contractual penalty clauses (\emph{Vertragsstrafen}) are not decoration
in this theory but the mechanism by which provability risk is priced
back into the signal. The threshold part is driven entirely by the
fixed-cost block \(c_0\): recoveries scale with the ticket while access
costs do not, so there is a project size below which no rational client
would ever enforce, every low type knows it, and the warranty means
nothing. At the stylized German-fee-schedule parameters
(\(c_0 = \text{€}2{,}000\), \(c_1 = 7.9\), \(\alpha = 0.66\),
\(\delta = 2\)), \(v_{\min}\) runs from roughly €450 at \(\rho = 0.75\)
to roughly €1,450 at \(\rho = 0.55\)---and these are \emph{lower
bounds}, since the unmodeled costs of suing (management time,
relationship destruction, the reputational cost of being a plaintiff)
shift the effective threshold upward, plausibly by an order of
magnitude. What matters is the structure, not the euro figure:
\textbf{the fixed-cost block of civil procedure removes the last costly
signal exactly where tickets are small---which is exactly where
AI-enabled entry is easiest.} Arbitration clauses with capped fees,
buyer-side legal-expenses insurance, and the bundling of small
engagements into framework contracts whose aggregate ticket clears
\(v_{\min}\) are, in this light, signaling infrastructure rather than
legal plumbing.

\hypertarget{insurers-judgment-proofness-and-the-retained-risk-principle}{%
\subsection{Insurers: judgment-proofness and the retained-risk
principle}\label{insurers-judgment-proofness-and-the-retained-risk-principle}}

The promise to pay \(D\) is also bounded by the ability to pay it. An
uninsured seller with wealth \(S\) can credibly promise at most
\(D_{\mathrm{eff}} = \min\{D, S\}\)---Shavell's (1986) judgment-proof
problem---which for small expert firms binds long before the
\(\varphi D \geq v\) condition is met on large engagements. Professional
indemnity insurance (\emph{Vermögensschadenhaftpflicht}) raises capacity
to \(\min\{D,\, S + I\}\). But insurance re-allocates precisely the cost
that made the promise informative: if the insurer pays the damages, the
mimic's imitation cost is whatever the \emph{insurer} charges him plus
whatever the contract makes him retain.

Formally, let cover pay verified damages above a deductible \(SB\), with
\(0 \leq SB \leq D\) throughout (for \(SB > D\) the retained amount is
\(\min\{SB, D\}\) and the analysis applies to it), at premium \(\pi\),
with standard exclusions for intent and knowing breach. A
type-\(\theta\) seller's outcome-contingent cost becomes
\(\pi_\theta + (1 - q_\theta)\,\varphi\, SB\). Two polar premium regimes
bracket reality:

\textbf{Proposition 7 (Retained risk).} \emph{(i) Under type-blind
pricing at the equilibrium book (``pooled'' hereafter)---cover priced by
a competitive, zero-loading insurer at the book it actually insures, so
that in the candidate separating equilibrium, where only high types buy,
\(\pi = (1 - q_H)\,\varphi\,(D - SB)\), identical for every insured---a
warranty supports separation if and only if the seller's actually borne,
collectible retention \(R = \min\{SB,\, S\}\) satisfies
\(\varphi R \geq v\) (strictly deterring for \(\varphi R > v\), the
mimic indifferent at equality)---nominal figures do not carry the
incentive constraint once wealth binds, so feasibility requires
\(D \geq SB\) with \(R \geq v/\varphi\), the total collectible transfer
capped at
\(D_{\mathrm{eff}} = \min\{D,\, S + I,\, D_{\mathrm{legal}}\}\)---: only
retained, collectible exposure carries information. Absent this pricing
tie, the premium wedge itself enters the deterrence condition, and
``pooled'' alone pins nothing down; \(SB\) is a fixed, observable
contract term here---were the deductible a hidden seller choice priced
at the high-type book, the boundary \(\varphi SB = v\) would be locally
fragile, and strictness \(\varphi SB > v\), or beliefs conditioning on
\(SB\), restores it. (ii) Under perfect actuarial type pricing---the
informational limit of experience rating,
\(\pi_\theta = (1 - q_\theta)\,\varphi\,(D - SB)\)---separation obtains
if and only if \(\varphi (R + C) \geq v\), where \(C\) is the
insurer-funded collectible cover (\(C \leq \min\{D - SB,\, I\}\),
\(R + C \leq D_{\mathrm{eff}}\))---for fully collectible contracts
\(R + C = D\) and the original \(\varphi D \geq v\) reappears: the
premium channel prices the insured part type-differentially; empirical
experience rating is noisy, dynamic convergence toward this limit
(Section 5.6). In both regimes the signal-effective liability is the
risk the seller retains or is individually priced on: \(R\) under
pooling and \(R + C\) under full rating, with intermediate
credibility-weighted regimes in between.}

The corollary deserves plain language. Under experience rating the
insurer prices \((1 - q_\theta)\)---the exact quantity the buyer cannot
observe---and thereby becomes a \textbf{delegated competence assessor}:
a certificate of cover with a high deductible, or with a genuinely
risk-rated premium, is informative about the insured---provided the
deductible or the rated premium is observable to the buyer or
contractually consequential; a certificate with a pooled premium and a
token deductible is not a signal at all but reinsurance of the mimic,
and in the aggregate it is the insurer's balance sheet, not the market's
information, that absorbs the imitation. The assessor role presupposes
risk adjustment: because abler sellers may attract harder engagements
(Section 3.1), raw claim frequencies conflate case mix with competence,
and informative rating conditions on engagement type. Rothschild and
Stiglitz's (1976) screening logic points in this direction---deductible
menus can sort risks---though that same literature warns that
competitive screening equilibria may fail to exist; we state the
delegated-assessor role as a mechanism hypothesis---for regime (i) as
for (ii)---illustrated rather than proven by the simulation; Section
5.6's ``pooled premium'' prices the realized mixed book and thus
illustrates the spirit of Proposition 7(i), not its candidate-book
pricing. Experience rating, moreover, requires claims \emph{histories},
and histories take time to accumulate. Section 5.6 quantifies this lag,
which turns out to be first-order: after an AI shock, the speed at which
liability-insurance premiums become risk-differentiated governs how long
the expert market stays pooled.

\textbf{Coverage boundaries in practice.} One drafting detail of
real-world policies sharpens the proposition rather than qualifying it.
Standard market wordings of German professional-indemnity insurance for
pure financial loss (\emph{Vermögensschaden-Haftpflicht}) insure
\emph{statutory} liability only: the classical exclusion---§ 4 no. 2 of
the traditional AVB model wording, found in many (not all: the model
terms are non-binding and individual wordings vary) current insurer
conditions---removes claims insofar as they exceed the scope of the
seller's statutory liability by virtue of contract or special promise
(AVB-V, § 4 no. 2; Austrian standard wordings contain a materially
identical clause, and Austrian case law consistently holds that
liability cover does not extend to the insured's pure performance
interest---OGH, Rechtssatz RS0081898). A warranty of the kind modeled
here---damages \(D\) promised beyond what statute would
award---therefore falls \emph{outside} default cover unless the insurer
endorses it individually. Two consequences follow inside the model.
First, by default the supra-statutory component of the promise rides on
the seller's own balance sheet, so the judgment-proof cap
\(D_{\mathrm{eff}} = \min\{D, S\}\) binds on exactly that
component---and the retained-risk principle holds \emph{a fortiori}: the
excess cannot be pooled at all, which makes it the purest form of the
signal---and the narrowest: combining Propositions 6 and 7, the
uninsured-excess route exists only in the window
\(v \in (v_{\min}(\rho),\, \hat{R}/\delta)\), with \(\hat{R}\) the
verified collectible retention---at \(\hat{R} = €250{,}000\) and
\(\delta = 2\), roughly €1,100 to €125,000. For this paper's own target
population, firms of 10 to 250 employees, the route closes exactly where
stakes are largest; the escrow, bond, and endorsement instruments of
this section are then not alternatives but the only carriers. Second,
where insurers do write such endorsements, they underwrite the
individual promise---Proposition 7's delegated-assessor channel
operating through contract design rather than premium history; the
affirmative AI-specific errors-and-omissions products noted in Section
7.3 are this channel in motion. The same logic runs up the value chain:
where the seller holds recourse against an upstream provider (a model
vendor, a subcontractor), recourse is economically one more layer of
cover, and Proposition 7 applies mutatis mutandis---the signal-effective
quantity is the risk that remains with the seller after all cover
\emph{and} recourse; liability that merely circulates along the value
chain informs the buyer of nothing. The legal discussion throughout
supplies examples consistent with the reduced form actually modeled;
enforceability, moderation, standard-terms control, cost allocation,
settlement, collectibility, and policy exclusions remain contract- and
jurisdiction-specific. For regulated professions, profession-specific
rules on liability limitation and remuneration add further constraints
on contract design, which we flag as scope conditions rather than model.

\hypertarget{simulation-study}{%
\section{Simulation Study}\label{simulation-study}}

The propositions are statements about equilibria of an idealized
continuum market. To probe whether they survive finite populations,
adaptive behavior, sampling noise in outcomes, and learning by would-be
mimics, we implement the model as an agent-based Monte-Carlo simulation.
We emphasize its epistemic status: the simulation \textbf{illustrates
and stress-tests the model's internal logic; it is not empirical
evidence}. Parameters are stylized; where we mention empirical
magnitudes alongside simulated ones, they serve as orientation, not
calibration targets.

\hypertarget{design}{%
\subsection{Design}\label{design}}

A market of \(N_H = 600\) high-type and \(N_L = 600\) low-type sellers
runs for \(T = 96\) months over \(R = 500\) replications per regime. The
AI frontier follows a logistic path \(a_t\) from \(0.30\) to \(0.97\)
with midpoint at month 36 (the ``AI shock'') and speed \(\tau = 4\).
High types draw outside options \(r \sim U[0.40,\, 0.80]\) (in units of
\(v\)) and participate when their smoothed expected payoff covers \(r\)
(exponential smoothing 0.7, allowing exit and re-entry). The simulated
support extends below \(q_L v = 0.55\) and thus relaxes the restriction
\(\underline{r} > q_L v\) under which the \(\lambda = 0\) clause of
Proposition 2 and part (i) of Proposition 5 are stated: with
\(F(q_L v) > 0\), complete unraveling to \(\lambda = 0\) is excluded by
construction. The relaxation is conservative for the collapse
result---the simulation understates, rather than overstates, the decline
the theory permits---and no reported number invokes the restriction.
Outcomes are drawn binomially each period; failures under warranty are
verified with probability \(\varphi\) and damages are actually paid, so
all payoffs are realized, not expected. Low types experiment with
warranty mimicry at an initial rate of 5\% (\(\varepsilon\)-greedy) and
reduce experimentation as realized losses accumulate. Table 2 lists all
parameters. Two regimes are compared: \textbf{Regime A} (production-side
signals only) and \textbf{Regime B} (liability contracts additionally
available, \(D = 1.25\,v\), \(\varphi = 0.9\), so
\(\varphi D = 1.125\,v \geq v\), satisfying Proposition 3's condition;
the minimum would be \(D_{\min} = v/\varphi \approx 1.11\,v\)).

\begin{longtable}[]{@{}
  >{\raggedright\arraybackslash}p{(\columnwidth - 4\tabcolsep) * \real{0.3333}}
  >{\raggedright\arraybackslash}p{(\columnwidth - 4\tabcolsep) * \real{0.3333}}
  >{\raggedright\arraybackslash}p{(\columnwidth - 4\tabcolsep) * \real{0.3333}}@{}}
\caption{Baseline simulation parameters. All monetary quantities in
units of \(v\). Parameters are stylized; the simulation is an
illustration of the model, not an estimate of any
market.}\tabularnewline
\toprule\noalign{}
\begin{minipage}[b]{\linewidth}\raggedright
Parameter
\end{minipage} & \begin{minipage}[b]{\linewidth}\raggedright
Symbol
\end{minipage} & \begin{minipage}[b]{\linewidth}\raggedright
Value
\end{minipage} \\
\midrule\noalign{}
\endfirsthead
\toprule\noalign{}
\begin{minipage}[b]{\linewidth}\raggedright
Parameter
\end{minipage} & \begin{minipage}[b]{\linewidth}\raggedright
Symbol
\end{minipage} & \begin{minipage}[b]{\linewidth}\raggedright
Value
\end{minipage} \\
\midrule\noalign{}
\endhead
\bottomrule\noalign{}
\endlastfoot
Buyer value of solved problem (numéraire) & \(v\) & 1.00 \\
Success probability, high / low type & \(q_H,\ q_L\) & 0.85, 0.55 \\
Artifact cost slope, high / low type & \(k_H,\ k_L\) & 1.0, 3.0 \\
Buyer discernment ceiling & \(\bar{s}\) & 1.00 \\
AI frontier, pre / post shock & \(a_{\text{pre}},\ a_{\text{post}}\) &
0.30, 0.97 \\
Shock midpoint, speed & \(t_0,\ \tau\) & month 36, 4.0 \\
Implied thresholds & \(\eta^{*},\ D_{\min}\) & 0.10, 1.11 \\
Warranty damages, verifiability & \(D,\ \varphi\) & 1.25, 0.90 \\
Sellers (high / low) & \(N_H,\ N_L\) & 600, 600 \\
High-type outside options & \(r\) & \(\sim U[0.40, 0.80]\) \\
Months, replications & \(T,\ R\) & 96, 500 \\
\end{longtable}

\hypertarget{dynamics-collapse-versus-preservation}{%
\subsection{Dynamics: collapse versus
preservation}\label{dynamics-collapse-versus-preservation}}

Figure 2 shows the central result. In Regime A, the market reproduces
Proposition 1 and 2 dynamically: while \(\eta_t \geq \eta^{*}\), high
types separate through artifacts and earn a net payoff of \(0.75\,v\)
(the separating price \(0.85\,v\) minus signaling cost \(0.10\,v\));
when the frontier passes \(a_t > 0.90\)---nine months after the shock
midpoint, in month 45---separation becomes infeasible, the premium
collapses to zero, and payoffs converge to the pooled price. In the
imposed pooling benchmark the market breaks at the threshold crossing
\(\eta_t < \eta^{*}\), not at the capability curve's inflection point.
The sharpness is benchmark-conditional: Proposition 1(ii)'s tight
envelope permits partially separating play paying up to
\(q_L v + \eta(k_L - k_H)\), which dominates the pooling fixed point for
\(\eta > 0.061\)---months 45--48 on this steep path, a four-month glide
rather than a break, lengthening under a slower \(a(A)\). The reading
``market breaks trail capability milestones'' is therefore a property of
the benchmark selection, not of the model, and the two are empirically
distinguishable. High-type participation falls from 88.1\% to 68.0\%
(averages over months 1--24 vs.~73--96 across 500 runs), the
endogenous-composition fixed point of Proposition 2
(\(\lambda^{*} \approx 0.40\), \(\bar{p} \approx 0.67\,v\)). Average
transacted quality erodes from 0.690 to 0.671---a modest 2.7\%: the
simulation shows \emph{partial}, not complete, unraveling under our
outside-option distribution, and we flag that heavier-tailed outside
options produce correspondingly deeper unraveling. The dramatic effect
is distributional: the high type's net payoff falls by 10.5\% while the
identification premium---the difference between what an identified
expert and the pool commands---falls from \(0.20\,v\) to zero. That
pattern---premium compression concentrated on the
previously-separated---is qualitatively the shape of the evidence in Hui
et al.~(2024), where top-rated freelancers lost disproportionately.

In Regime B, nothing of the sort happens: high types adopt the warranty,
are identified, and post \(p_H = 1.019\,v\); their realized net payoff
averages \(0.850\,v\) (the theoretical \(q_H v\), damages actuarially
fair at \(N = 600\)); participation is essentially complete throughout
(99.8\% pre-shock average, 100\% post-shock); the net premium holds at
\(0.300\,v\) pre- and post-shock. Low-type mimicry is
self-extinguishing: experimentation falls from 5\% to 0.7\% by month 96
as realized damages teach mimics that \(\varphi D \geq v\) makes
imitation a losing trade. The information the market needs survives the
shock entirely---carried by the one signal whose cost the shock could
not touch.

\begin{figure}
\centering
\includegraphics[width=1\textwidth,height=\textheight]{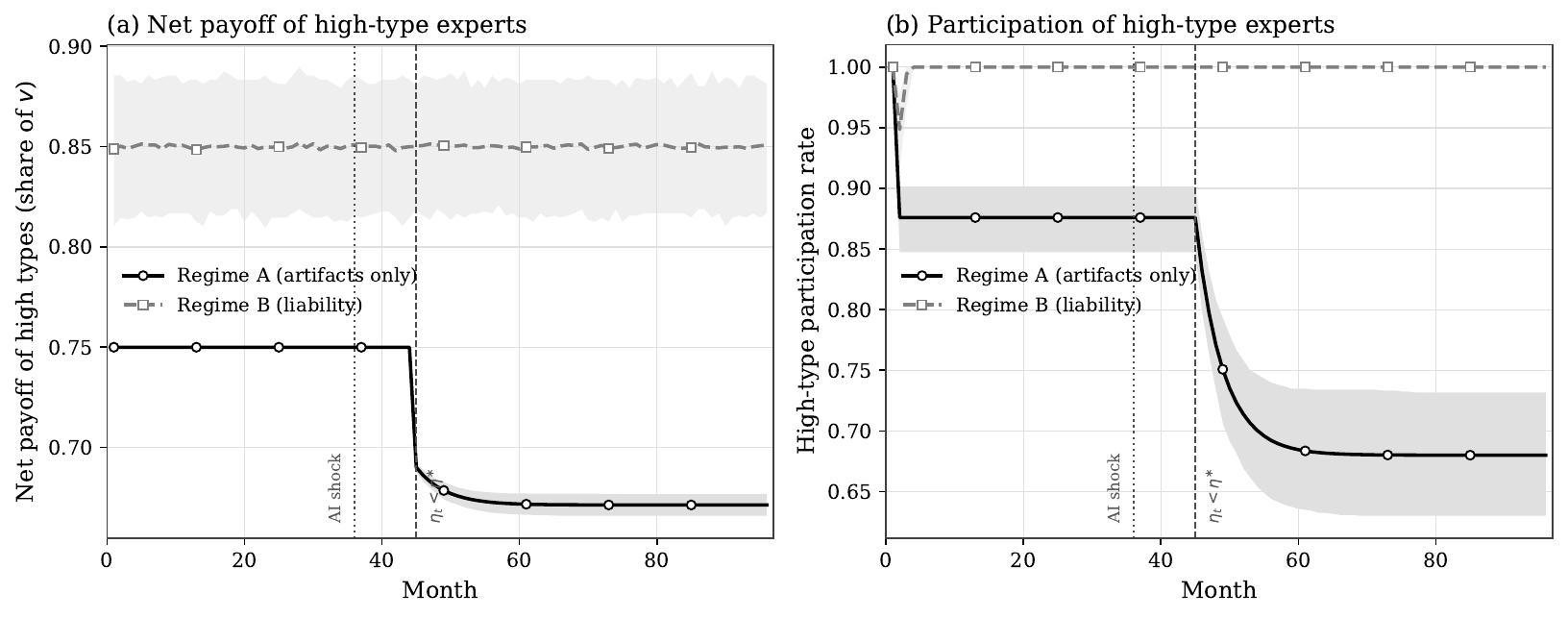}
\caption{Market dynamics around the AI capability shock, 500 Monte-Carlo
replications; shaded bands are 2.5--97.5\% ranges. The dotted vertical
line marks the shock midpoint (\(t_0 = 36\)); the dashed vertical line
marks the month in which the headroom crosses the threshold
(\(\eta_t < \eta^{*}\), month 45)---the market breaks at the threshold
crossing, not at the shock midpoint. (a) Net payoff of high-type
experts. (b) Participation of high-type experts. Solid: Regime A
(production-side signals only) --- the premium collapses and roughly a
fifth of high types exit. Dashed: Regime B (liability contracts
available) --- separation, payoff, and participation are fully sustained
(Proposition 3).}
\end{figure}

\hypertarget{the-phase-map}{%
\subsection{The phase map}\label{the-phase-map}}

Figure 3 locates the two thresholds in the \((\varphi D / v,\ \eta)\)
plane. The analytic boundaries \(\eta = \eta^{*}\) (Proposition 1) and
\(\varphi D = v\) (Proposition 3) partition the plane into: (i) artifact
separation feasible (upper region), (ii) liability separation (right
region), and (iii) pooling/unraveling (lower-left). Overlaid dots show
the fraction of reduced-form finite-sample runs in which separation is
sustained at each grid point; the simulated frequencies respect the
analytic boundaries, with the expected sampling blur close to
\(\varphi D = v\). Generative AI moves markets \emph{downward} in this
map; the only exit to the right is contractual.

\begin{figure}
\centering
\includegraphics[width=0.72\textwidth,height=\textheight]{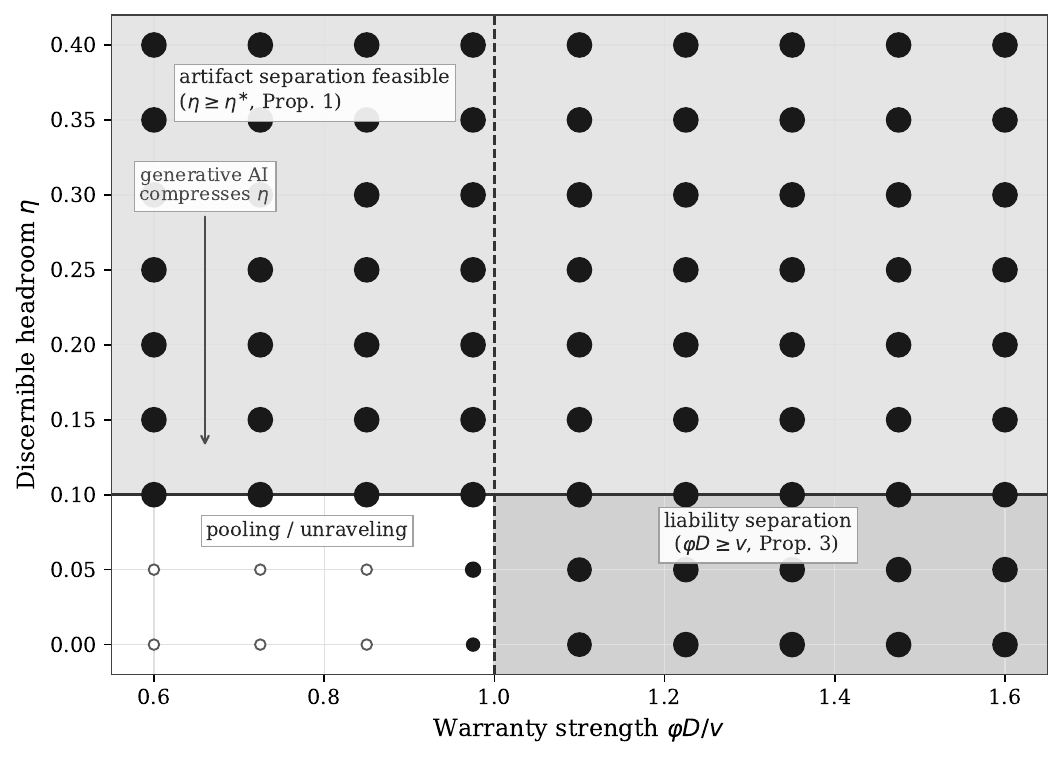}
\caption{Phase map of separating equilibria. Shaded regions: analytic
(Propositions 1 and 3). Dots: fraction of finite-sample simulation runs
sustaining separation (dot area proportional to frequency; open dots
\(\approx\) 0). AI capability compresses \(\eta\) (vertical descent);
warranty strength \(\varphi D\) is a design choice (horizontal
position).}
\end{figure}

\hypertarget{the-rising-blade-in-finite-samples}{%
\subsection{The rising blade in finite
samples}\label{the-rising-blade-in-finite-samples}}

Figure 4 verifies Proposition 4's comparative static: the simulated
return to being identified tracks the analytic line
\((1 - \lambda)\,\Delta q\, v\) across the full range of pool
compositions, with finite-sample dispersion. As AI-enabled entry pushes
\(\lambda\) down, the value of the surviving signal rises
mechanically---no change in preferences, no anti-AI sentiment, pure
composition. This matters for interpretation: the model predicts a
rising liability premium \emph{even in a world where nobody minds
machines}, which distinguishes it from novelty-aversion accounts and
gives Section 7's experiment one of its discriminating hypotheses.

\begin{figure}
\centering
\includegraphics[width=0.8\textwidth,height=\textheight]{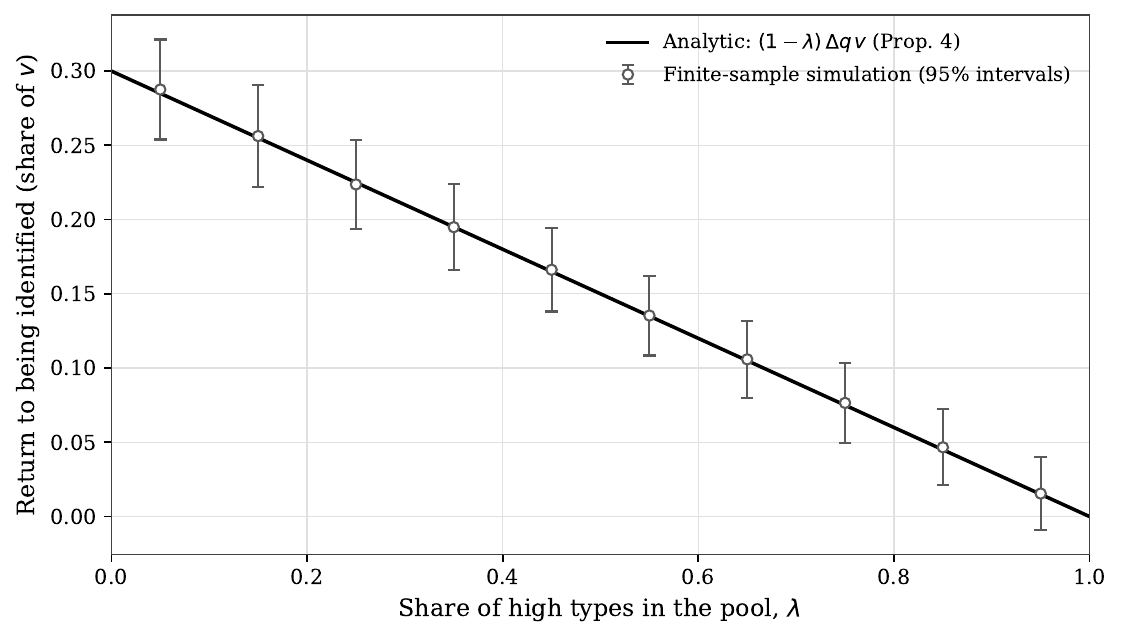}
\caption{Return to being identified as high-type, as a function of the
share \(\lambda\) of high-quality providers in the pool. Line: analytic,
\((1-\lambda)\Delta q v\) (Proposition 4). Dots: finite-sample
simulation (800 sellers per draw, 200 draws per point; bars are 95\%
intervals).}
\end{figure}

\hypertarget{the-small-ticket-exclusion-in-the-agent-based-market}{%
\subsection{The small-ticket exclusion in the agent-based
market}\label{the-small-ticket-exclusion-in-the-agent-based-market}}

To probe Proposition 6 with heterogeneous provability, each
seller--buyer match draws
\(\rho \sim \mathrm{Beta}(24\rho_0,\, 24(1-\rho_0))\), damages scale as
\(D = 2v\), and ticket sizes run over a logarithmic grid from €200 to
€300,000 (200 replications of 400 matches per grid point;
litigation-cost parameters as in Section 4.1). Figure 5 shows the
result: the share of matches in which the liability signal is viable
rises in a sharp S-curve through the analytic
threshold---\(v_{\min} = \text{€}1{,}086\) at \(\rho_0 = 0.60\),
\(\text{€}451\) at \(\rho_0 = 0.75\)---and is essentially zero below
€500. Two features merit note. First, the transition is smooth rather
than a step only because provability is heterogeneous; the analytic
threshold sits at the curve's midpoint, as it should. Second, the
\(\rho_0 = 0.60\) curve plateaus near 0.84, not 1: roughly 15\% of
matches draw \(\rho < 1/\delta = 0.5\), for which Proposition 6's
deterrence condition fails \emph{at every ticket size}. Provability, not
project size, is the binding constraint for that residual---a reminder
that ex-ante agreed acceptance criteria, which raise \(\rho\), are as
much a part of the signal as the damages clause itself.

\begin{figure}
\centering
\includegraphics[width=0.88\textwidth,height=\textheight]{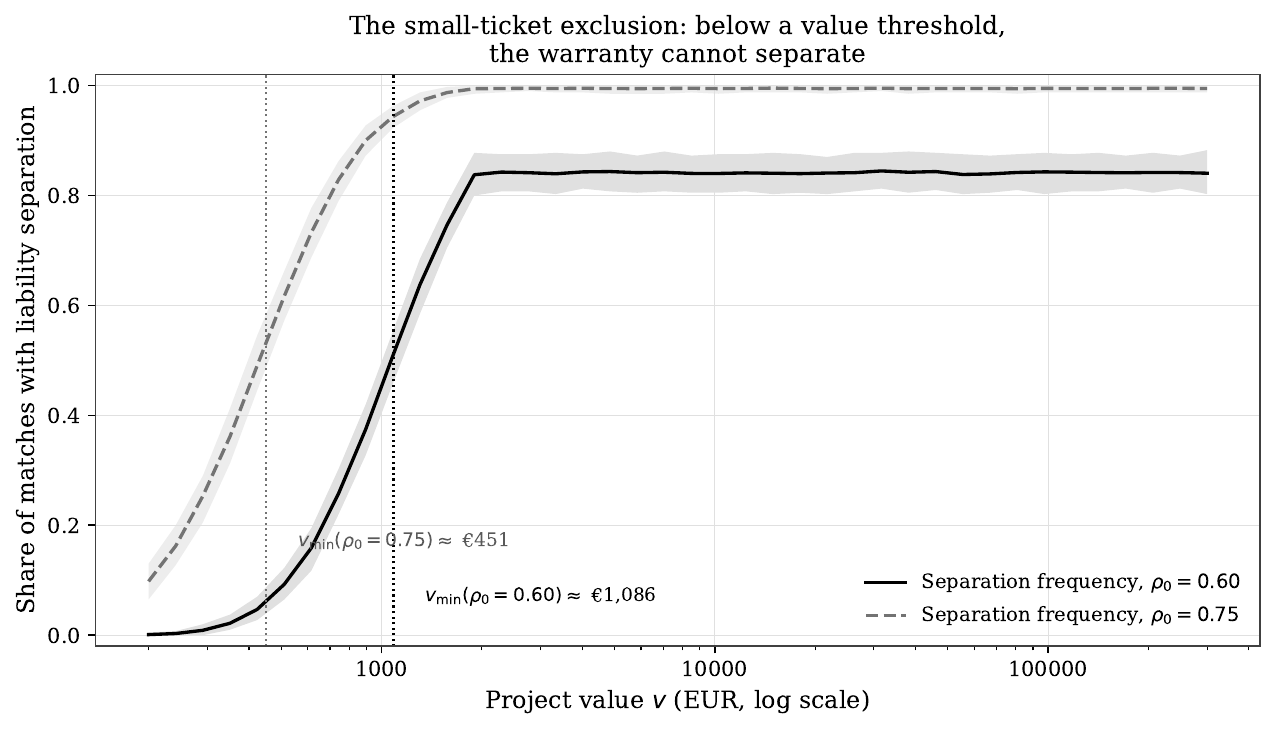}
\caption{Viability of the liability signal by project value (EUR, log
scale). Lines: mean share of matches satisfying credibility and
deterrence (Proposition 6) at \(\rho_0 = 0.60\) (solid, with 95\% band)
and \(\rho_0 = 0.75\) (dashed); dotted verticals: analytic \(v_{\min}\).
Litigation-cost parameters stylized on German fee-schedule magnitudes
(Section 4.1).}
\end{figure}

\hypertarget{insurance-regimes-in-the-agent-based-market}{%
\subsection{Insurance regimes in the agent-based
market}\label{insurance-regimes-in-the-agent-based-market}}

To probe Proposition 7, warranted sellers carry professional indemnity
cover (\(D = 2v\), \(\varphi = 0.9\)) under three regimes: pooled
premium with a low deductible (\(SB = 0.3v\), so
\(\varphi SB = 0.27 < v\)), pooled premium with a high deductible
(\(SB = 1.4v\), so \(\varphi SB = 1.26 \geq v\)), and experience rating
with the low deductible (credibility-weighted individual premiums,
credibility horizon \(k = 20\) exposures). Low types adopt the warranty
adaptively (logistic choice on smoothed realized profit); insurers
charge \((D - SB)\) times the verified-claim frequency they
observe---pooled or individual---so that premiums equal Proposition 7's
actuarial benchmark \(\pi_\theta = (1 - q_\theta)\varphi(D - SB)\) in
expectation; buyers update the belief \(P(H \mid \text{warranty})\) from
the realized success rate of warranted engagements (200 replications).

Figure 6 shows the three futures. Under the pooled low-deductible
regime, mimicry is a subsidized trade: low-type adoption rises to 99\%,
the posterior collapses to the prior 0.50 within months, and the
warranty carries no information---with the losses landing on the
insurer's book rather than on the mimics, exactly the configuration
Proposition 7(i) predicts. Under the pooled high-deductible regime the
market barely notices the temptation: mimicry stays in single digits (an
exploration floor) and the posterior holds near 0.95. The
experience-rated regime is the instructive one: it collapses
\emph{first}---the posterior reaches its trough of 0.50 by month 8,
while new mimics still pay near-pooled premiums---and then recovers as
claims histories accumulate: mimicry falls below half its peak by month
30 and to about 2\% in the final year, and the posterior climbs back to
0.98 (final-year average). Two panels of Figure 6 display the machinery
behind this reversal. Panel (c) shows the insurer acting as the
delegated competence assessor of Proposition 7: premiums charged to high
types converge to actuarial cost (0.231 in the final year, against the
fair benchmark \((1-q_H)\varphi(D-SB) = 0.230\)), while the premium
charged to active mimics rises far above the pooled rate---but stalls
near 0.51, well short of the low type's actuarial cost of 0.69, because
the mimic pool continually refreshes with newcomers whose thin claims
histories still command near-pooled rates: experience rating disciplines
the \emph{stock} of mimics while subsidizing the \emph{flow}. Panel (d)
adds a second clock: buyers' belief \(\mu_t\), updated from realized
success rates, lags the true posterior by roughly a year on the recovery
path. The rating channel works, but on the insurer's---and then the
buyer's---information timescale, not the market's. This is a result we
did not anticipate when specifying the model, and it sharpens the policy
reading of Proposition 7: \textbf{after an AI shock, the speed at which
liability-insurance premiums become risk-differentiated is a first-order
determinant of how long the expert market stays pooled.}

\begin{figure}
\centering
\includegraphics[width=1\textwidth,height=\textheight]{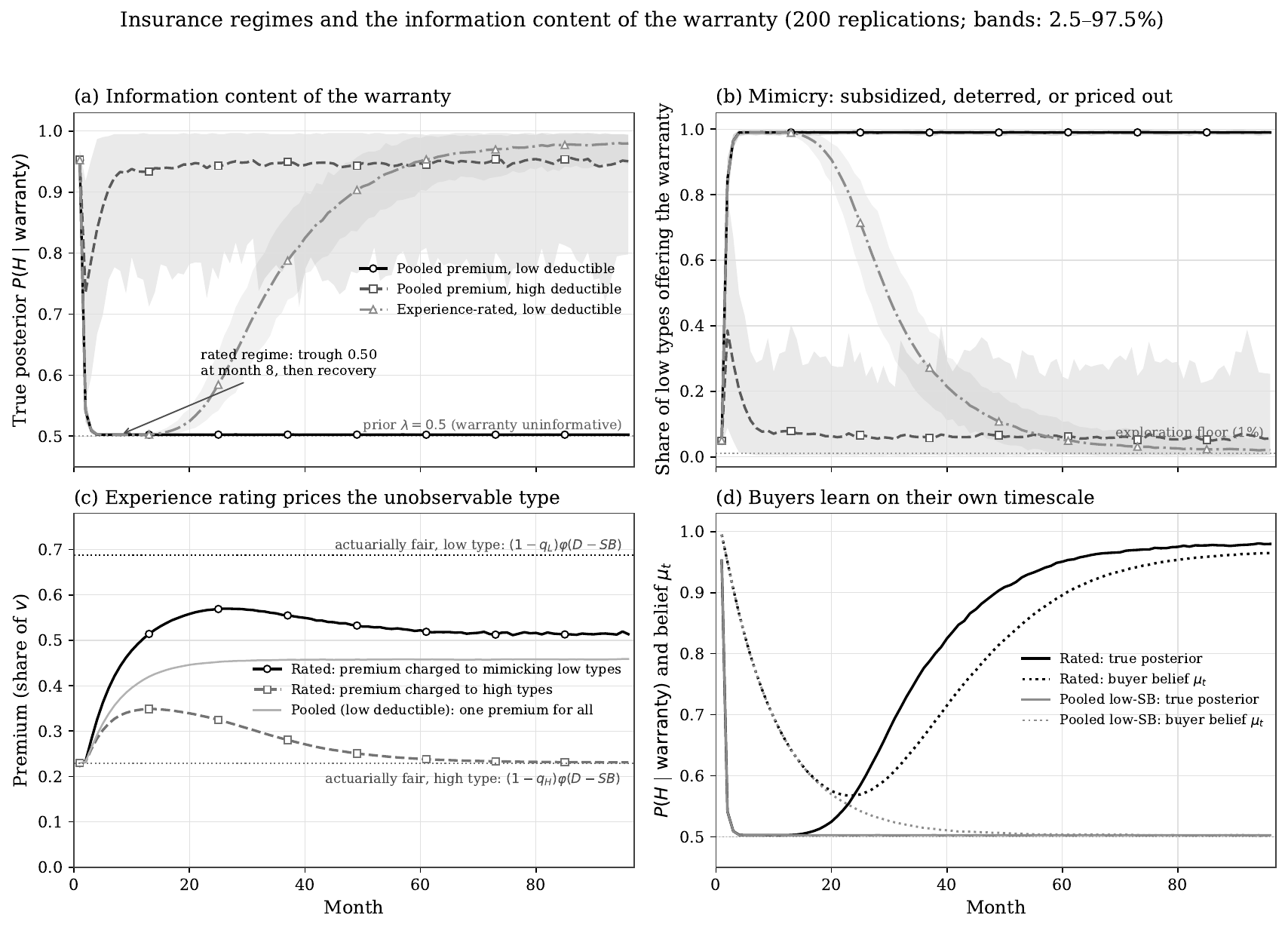}
\caption{Insurance regimes and the information content of the warranty
(200 replications; bands are 2.5--97.5\% ranges). (a) True posterior
\(P(H \mid \text{warranty})\): pooled premium with low deductible
(solid) renders the warranty uninformative; pooled premium with high
deductible (dashed) sustains separation (Proposition 7(i),
\(\varphi SB \geq v\)); experience rating with low deductible
(dash-dotted) collapses first, then recovers as claims histories
accumulate (Proposition 7(ii)). (b) Share of low types offering the
warranty: subsidized under pooled-low, deterred under pooled-high,
priced out---slowly---under rating. (c) The delegated-assessor
mechanism: under experience rating, premiums charged to high types
converge to actuarial cost, while premiums charged to active mimics rise
toward, but never reach, the low type's actuarial cost---new mimics with
thin histories keep entering at near-pooled rates. (d) Buyers' belief
\(\mu_t\) versus the true posterior: buyers learn from realized success
rates and lag the truth in both directions.}
\end{figure}

\hypertarget{the-narrative-in-one-figure}{%
\subsection{The narrative in one
figure}\label{the-narrative-in-one-figure}}

Figure 7 compresses the simulation study into its four-step narrative,
overlaying regimes within each panel so that the institutional
comparison---not the level of any single series---carries the
information. (a) The competence premium survives the shock midpoint by
nine months and then vanishes within months under artifacts only, while
under liability it is untouched. (b) Roughly a fifth of high types exit
in Regime A; none do in Regime B. (c) Average delivered quality erodes
by 2.7\% in Regime A and holds at the full-participation mix
\((q_H + q_L)/2 = 0.70\) in Regime B. (d) The institutional epilogue
from Figure 6: whether the warranty stays informative under insurance
depends on the contract form, not on the existence of cover. Panels
(a)--(c) show the baseline dynamics around the AI shock; panel (d)
simulates the post-shock world from month 1, so no shock marker applies
there.

\begin{figure}
\centering
\includegraphics[width=1\textwidth,height=\textheight]{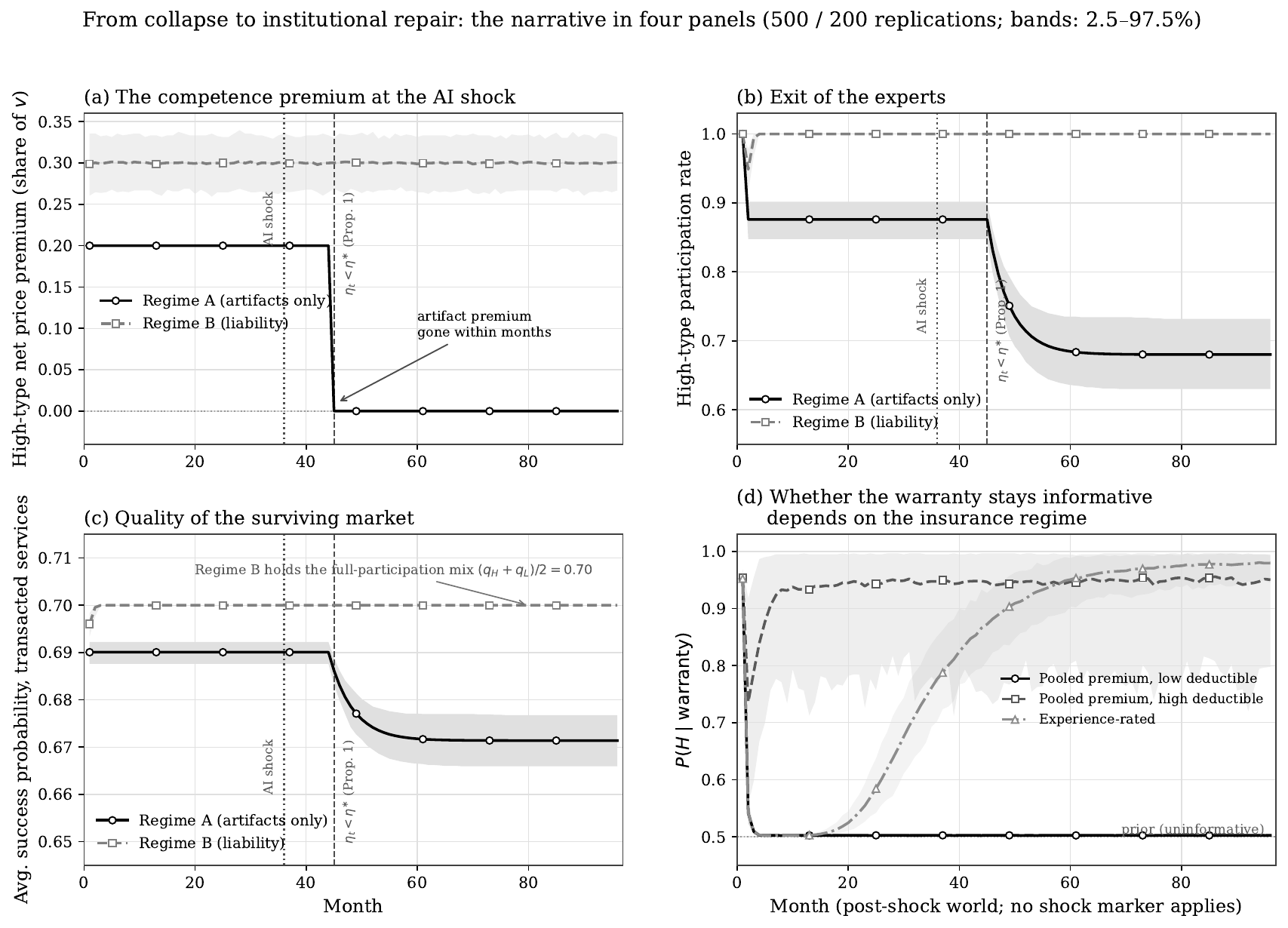}
\caption{From collapse to institutional repair. (a) High-type net price
premium, Regime A vs.~Regime B (dotted vertical: shock midpoint; dashed
vertical: \(\eta_t < \eta^{*}\), Proposition 1). (b) High-type
participation. (c) Average success probability of transacted services.
(d) \(P(H \mid \text{warranty})\) under the three insurance regimes of
Figure 6. Bands are 2.5--97.5\% ranges (500 replications in (a)--(c),
200 in (d)).}
\end{figure}

\hypertarget{what-the-simulation-does-not-show}{%
\subsection{What the simulation does not
show}\label{what-the-simulation-does-not-show}}

Four deliberate limitations. First, no parameter of Table 2 is
estimated; the simulation cannot confirm the theory against the world,
only against itself. Second, buyers in the model price competitively and
update correctly; behavioral pricing frictions (anchoring on historical
rates, procurement rules) are absent and would plausibly slow both the
collapse and the recovery. Third, all Regime-B series are drawn at
litigation rate \(\ell = 0\), the settlement benchmark: the seller-side
litigation wedge of Section 4.1 is absent, and at \(v = €5{,}000\) under
full litigation it would pull Regime-B participation to roughly
\(72\%\)---close to the artifact-collapse level, which is why the wedge
is a threshold result and not a footnote. Fourth, the baseline dynamics
treat \(\varphi\) and \(D\) as exogenous contract technology; Sections 4
and 5.4--5.5 endogenize both through stylized institutions, but the
institutional parameters---fee schedules, provability distributions,
credibility horizons---are themselves stylized rather than estimated.
The caveat moves one level down; it does not disappear.

\hypertarget{discussion}{%
\section{Discussion}\label{discussion}}

\hypertarget{theoretical-implications}{%
\subsection{Theoretical implications}\label{theoretical-implications}}

The paper's contribution to theory is deliberately narrow and, we hope,
sharp: a \emph{classification with a survival result}. Signals sort by
the locus of their cost---production-side versus
outcome-contingent---and a technology shock that deflates production
costs sorts them into casualties and survivors along exactly that line.
The mechanism is Spence's; the classification refines Kirmani and Rao's
(2000) default-independent/default-contingent distinction into a
comparative static in AI capability; the survival result (Proposition 3)
gives the credence-goods liability literature (Dulleck et al., 2011) a
signaling-side twin. The scissors corollary (Proposition 4) may be the
most useful piece for empirical work, because it converts a vague
cultural claim---``authenticity is becoming more valuable''---into a
composition effect with a sign, a magnitude, and no reliance on anyone's
attitude toward machines.

The provenance result (Proposition 5) speaks to an active policy
conversation. Mandatory AI labeling (EU AI Act, Art. 50) and content
credentials (C2PA) are frequently discussed as trust infrastructure. In
our framework they are \emph{origin} infrastructure: valuable for what
they do (tamper-evidence, attribution), structurally unable to do what
is sometimes hoped of them (restore competence signaling), for the
simple reason that their cost does not covary with competence. In
practice the case is even weaker than the proposition suggests:
provenance metadata is routinely stripped in real workflows---but our
point is that even \emph{perfect} provenance would not close the
information gap that matters here.

\hypertarget{what-ai-does-to-competence-itself}{%
\subsection{What AI does to competence
itself}\label{what-ai-does-to-competence-itself}}

The model holds \(q_\theta\) fixed as \(a\) rises: generative tools
change what documents cost, not what problems get solved. Three
objections to this invariance deserve explicit answers, because each has
current behavioral evidence behind it.

\emph{AI as substance.} Suppose the tools raise competence
itself---\(q_\theta(a)\) increasing, the substance channel rather than
the artifact channel. The separation condition is unaffected:
\(\Delta q\) cancels from the deterrence algebra (the mimic's price gain
is \(\Delta q\, v\), his extra expected warranty cost
\(\Delta q\, \varphi D\)), so Proposition 3's \(\varphi D \geq v\) holds
for any \(q_L < q_H\) and along any path \(q_\theta(a)\). What changes
are magnitudes: both types' expected warranty costs fall as \(q_\theta\)
rises, posted warranty prices fall with them, and the identification
premium \((1 - \lambda)\,\Delta q(a)\, v\) shrinks if AI compresses the
competence gap itself. In the limit \(\Delta q \to 0\) the information
problem dissolves and there is nothing left to signal---and adoption
dies first: with any fixed contracting cost, warranties are offered only
while \((1-\lambda)\,\Delta q(a)\,v\) clears it, so the participation
margin closes before the incentive constraint does. Not a failure of the
model but its boundary, and empirically it is the Humlum--Vestergaard
scope condition of Section 6.5 seen from the other side: where AI
equalizes the ability to \emph{solve} rather than merely to
\emph{present}, no signaling analysis applies, and none is needed. A
companion paper endogenizes exactly this margin: the human fallback
capability that solves what the machine cannot is built by keeping staff
engaged and erodes otherwise, so the outcome-contingent commitment
certifies an endogenous, perishable asset rather than a fixed type
(Bauer, 2026a).

\emph{Two-sided moral hazard.} A warranty may reduce the \emph{client's}
effort---cooperation, data provision, implementation discipline---and
thereby lower the realized success probability, raising the warranty's
expected cost endogenously: the classic double-moral-hazard problem of
warranty provision (Cooper \& Ross, 1985). The instruments practice has
evolved for it are already objects in this model: a deductible makes the
client's carelessness partly his own cost (the \(SB\) of Proposition 7),
and ex-ante agreed acceptance criteria with documented cooperation
duties raise \(\rho\) and hence \(\varphi\) (Proposition 6) while
defining the promise so that client-caused failure is not a covered
failure. We treat client effort as folded into \(q_\theta\); unfolding
it adds realism on the contract-design margin without disturbing the
signaling margin, and it is the natural next extension for the journal
version.

\emph{Delegation and dishonesty.} Recent evidence shows that people
request unethical behavior more readily when they can delegate it to
machine agents, and that machine agents comply with fully dishonest
instructions far more often than human agents do (Köbis et al., 2025;
thirteen studies, over 8,000 participants). Read in our terms, this
raises the propensity of low types to \emph{attempt} mimicry---an entry
parameter, not an equilibrium object. In the simulation that propensity
is explicit (the exploration rate \(\varepsilon\) of Appendix B); under
\(\varphi D \geq v\) mimicry remains a losing trade in expectation, so a
higher \(\varepsilon\) produces more early attempts and therefore
faster, more visible losses---changing the speed, not the destination,
of Regime B's self-extinguishing dynamic---while in Regime A, where
artifacts are free to fake, it accelerates exactly the pooling the
theory predicts. The behavioral evidence thus tightens the paper's
motivation rather than loosening it: cheaper dishonesty is precisely the
environment in which outcome-priced signals earn their keep.

\emph{AI as the seller.} The strongest version of the objection makes AI
not the provider's tool but the provider: autonomous LLM agents
transacting in expert markets, with an objective function the buyer
cannot observe. Early evidence is directly on point. In one-shot
credence-goods experiments spanning human, machine, and mixed
(human--AI--human) markets, LLM experts extract substantially higher
surplus than human experts at consumers' expense, and a majority of
human experts delegates to such agents when allowed to set the agent's
objective function (Erlei, 2025, preprint). In agent-only markets
simulated with frontier models, one-shot interactions largely break
down---except under liability rules or explicitly pro-social agent
objectives---and the alignment of the agent's objective function emerges
as the first-order determinant of efficiency (Erlei \& Meub, 2026,
preprint): for machine agents, the institutional ranking that Dulleck et
al.~(2011) established for humans reappears, with liability again doing
the disciplining. In the model's terms, an exploitative objective
function is moral hazard layered on adverse selection---it lowers the
success probability the buyer actually experiences---and the remedy is
the same contingent claim: a \emph{disclosed} objective function is
process certification in the sense of Proposition 5, cheap to state and
cheap to change, whereas a liability contract prices the outcome
whatever the objective function was. That the last costly signal
disciplines sellers who are not human is not an extension of the theory;
nothing in \(\varphi D \geq v\) presupposes a human seller.

\hypertarget{managerial-implications}{%
\subsection{Managerial implications}\label{managerial-implications}}

For providers of expert services---consultancies, law and engineering
firms, agencies, specialized B2B services---the model reads as an
inventory instruction: list your signals, and ask of each one,
\emph{what does it cost to fake now?} Artifacts, polish, fluency, volume
of output: the model says their informational rent is going to zero, and
the empirical record (Hui et al., 2024) suggests the repricing is
underway. What remain expensive to fake are commitments whose cost is
contingent on outcomes: binding warranties, damages clauses, penalty
provisions---and their institutional carriers, from professional
indemnity insurance to licenses with revocation risk. Proposition 3's
condition \(\varphi D \geq v\) translates into two practical dials:
raise \(D\) (the bindingness of the promise---explicit acceptance
criteria, remediation at own cost, penalty clauses) and raise
\(\varphi\) (the verifiability of the outcome---defined success criteria
agreed ex ante, which \emph{reduce} the damages needed for the same
signal). A warranty over vague deliverables is a weak signal not because
promises don't work but because \(\varphi\) is near zero. One caution
from Section 4.2's coverage boundaries belongs in the same breath:
standard indemnity wordings insure \emph{statutory} liability only, so a
promise beyond it binds the firm's own balance sheet unless explicitly
endorsed---the policy should be read before the warranty is printed into
the offer. The model also warns against a tempting middle path:
certifying ``human-made'' as a marketing device. Proposition 5 says the
certificate itself carries no competence information; whatever premium
it commands rests on origin preferences of the Schilke--Reimann kind,
which the model treats as fragile relative to composition effects.

Proposition 6 adds a caveat with practical bite: none of this works
below the litigation threshold. Providers whose engagements are
individually small should not conclude that liability signaling is
closed to them, but that the unit of the promise must change---framework
contracts and annual retainers that aggregate small engagements above
\(v_{\min}\), arbitration clauses with capped fees that lower \(c_0\),
and ex-ante acceptance criteria that raise \(\rho\) are the instruments
that move a small-ticket practice back onto the viable side of Figure 5.

For buyers---procurement in particular---the model implies that the
cheapest quality screen available is to \emph{invite liability terms and
watch who walks away}. One legal asymmetry decides how the invitation
must be drafted. Provider-\emph{offered} warranties are largely safe
from standard-terms attack: the user of standard terms cannot invoke
§307 BGB against his own clause, and §348 HGB excludes judicial
moderation between merchants. Buyer-\emph{imposed} penalty clauses are
the mirror image: §307(1) BGB applies (§310(1) BGB preserves the control
between merchants), and the BGH caps standard-term contractual penalties
near 5\% of the settlement amount---judgment of 23 January 2003, VII ZR
210/01, tightened by judgment of 15 February 2024, VII ZR 42/22, which
struck even a 5\% cap keyed to the provisional contract sum. Against the
required \(D = v/\varphi \approx 1.11\,v\), an AGB-proof 5\% of the fee
is \(0.05\,v\) on this paper's pricing convention---a factor-22
shortfall, factor 222 at a realistic fee-to-value ratio---so the screen
works only as an \emph{invitation to offer}, with the provider as the
drafter; imposing the clause through procurement conditions destroys the
instrument it is meant to elicit, and individually negotiated terms
reintroduce Proposition 6's fixed-cost logic on the contracting side.
Under \(\varphi D \geq v\) the invited promise is nearly free for
competent providers in expectation and prohibitively priced for the
rest; self-selection does the assessment.

\hypertarget{policy-and-institutional-implications}{%
\subsection{Policy and institutional
implications}\label{policy-and-institutional-implications}}

Two institutions do heavy lifting in the model's background. First,
contract enforcement: \(\varphi\) and the credibility of \(D\) are
jurisdictional goods, and the model implies that the market value of
reliable, fast enforcement of service warranties \emph{rises} with AI
capability. Second, insurance: Section 4.2 develops the point in full.
Wealth constraints cap feasible \(D\) for small firms; indemnity
insurance restores capacity but relocates the signal into the retained
risk and the premium (Proposition 7), and the simulation shows the
transition path runs through a pooled valley whose length is set by how
fast insurers can rate risks individually. A market in which insurers
refuse cover, or will pool but not rate, is a market telling you its
information problem is unsolved---and indemnity premiums under
experience rating are an observable market price of \(1 - q_\theta\),
the quantity buyers cannot see. A third institution runs opposite to the
signal's scaling requirement and is the model's most falsifiable
institutional prediction: German professional law caps expert liability
in absolute euro amounts decoupled from any client's \(v\)---auditors
may limit liability in pre-formulated terms to four times the statutory
minimum cover and individually to the minimum itself (§54a WPO;
€4m/€1m), tax advisers analogously (§67a StBerG; €1m/€250k), lawyers
likewise (§51a BRAO). A signal that must scale with the value at risk
swims against a system of fixed ceilings calibrated to insurance minima;
the model then predicts observable strain exactly where AI exposure is
highest---renegotiated caps, carve-outs priced above the statutory
floor, migration of large-ticket engagements outside the capped
professions---each a live institutional test.

\hypertarget{boundary-conditions-and-honest-doubts}{%
\subsection{Boundary conditions and honest
doubts}\label{boundary-conditions-and-honest-doubts}}

Six limits deserve explicit statement. \textbf{(1) Scope of the falling
curve.} Humlum and Vestergaard's (2025) null result stands as a caution:
in occupations where value is produced in ways buyers can observe or
where artifacts are not the value proposition, the collapse mechanism
has nothing to bite on. Our claims are about credence-good expert
services, and aggregating them into claims about ``knowledge work''
would be refuted by existing data. \textbf{(2) The weight of liability
in real purchasing.} Von Wedel and Hagist (2022) found liability not
decisive in one professional purchasing context. If decision-makers
systematically undervalue liability relative to the model's predictions,
the separating equilibrium exists on paper and not in procurement. This
is precisely what Section 7's experiment measures rather than assumes.
\textbf{(3) Stated versus revealed preference.} Every willingness-to-pay
figure cited in this paper is stated, not transacted; the
stated--revealed gap in comparable settings is commonly estimated at a
factor of 1.5 to 2. Any premium the proposed experiment finds should be
reported with that discount attached. \textbf{(4) Transition, not
equilibrium.} The disclosure penalties documented by Schilke and Reimann
(2025) may partly reflect a transitional norm---early-days aversion that
habituation will erode, as it did for earlier technologies. The model is
deliberately built so that its central results do \emph{not} depend on
such aversion (Propositions 1, 3, and 4 contain no taste parameter over
machines), but the \emph{level} of any empirically measured premium
surely mixes composition effects with transitional sentiment, and only
panel evidence will separate them. \textbf{(6) Private diagnosis.}
Assumption (A1) switches off the credence good's defining feature.
Reinstate it---cheap AI triage is precisely a technology for private
task-level foresight---and a mimic warrants selectively: with selection
intensity \(q_{L,\mathrm{sel}} = E[q_L \mid L \text{ warrants}]\),
deterrence requires
\(D \geq (v/\varphi)\,\Delta q/(q_H - q_{L,\mathrm{sel}})\), the paper's
\(v/\varphi\) only without selection, diverging as
\(q_{L,\mathrm{sel}} \uparrow q_H\) (at a Table-2-consistent easy/hard
split: \(1.11\,v \to 3.33\,v \to \infty\)). Selection intensity is the
warranty's headroom analogue---the outcome-contingent counterpart of
\(\eta/\eta^{*}\)---so AI enters the liability signal after all, through
\(q_{L,\mathrm{sel}}(A)\) rather than through \(\kappa_\theta\);
bounding warranty informativeness by
\((q_H - q_{L,\mathrm{sel}})/\Delta q\) is the natural companion result,
left to the sequel. \textbf{(5) One mechanism, many markets.} The scope
is credence-good B2B expert services. The adjacent markets this paper
cites for evidence respond to disclosure and liability through different
channels: consumer and creative markets supply the disclosure-discount
evidence but trade origin as a taste attribute rather than a competence
signal; platform gig markets exhibit the falling curve with reputation
systems partly substituting for contract; regulated professions operate
under profession-specific liability and remuneration rules that
constrain contract design (Section 4.2). The mechanism travels; the
parameter values do not, and none of the quantitative claims here should
be exported across these market boundaries without re-derivation.

\hypertarget{the-proposed-experiment-and-falsification}{%
\section{The Proposed Experiment and
Falsification}\label{the-proposed-experiment-and-falsification}}

\hypertarget{design-1}{%
\subsection{Design}\label{design-1}}

The gap of Table 1 calls for a preregistered \textbf{choice-based
conjoint (CBC) experiment} with business decision-makers in the
German-speaking B2B expert-services market. Respondents---owners and
managing directors of firms with 10--250 employees, plus qualified
procurement leads---choose repeatedly among profiles of an otherwise
identical expert engagement (fixed scope description, e.g., a technical
due-diligence report or a compliance assessment), varied along five
attributes:

\begin{enumerate}
\def\labelenumi{\arabic{enumi}.}
\tightlist
\item
  \textbf{Warranty/liability} (3 levels): none stated / explicit
  acceptance criteria with remediation at provider's cost / remediation
  plus contractual penalty---with the engagement's value at risk stated
  in every vignette and the remediation and penalty levels quantified as
  \(0.5\times\), \(1\times\), and \(2\times\) that value, so the
  profiles span the threshold \(\varphi D = v\);
\item
  \textbf{Production disclosure} (3 levels): no statement / ``produced
  with substantial AI assistance'' / ``produced exclusively by named
  senior experts'';
\item
  \textbf{Price} (4 levels, bracketing prevailing daily-rate benchmarks,
  e.g., anchored to the BDU (2025) reported mean of €1,300);
\item
  \textbf{References} (2 levels: none / two named, contactable
  references) --- control attribute;
\item
  \textbf{Turnaround time} (2 levels) --- control attribute.
\end{enumerate}

Each respondent additionally states a subjective success probability
\(\hat{q}\) for the described engagement before the choice tasks, so
that willingness to pay can be benchmarked against the actuarial value
\((1-\hat{q})\varphi D\); and one limit is structural and stated up
front---randomized attributes cannot carry self-selection, the
respondent knowing the warranty was assigned rather than chosen, so the
conjoint identifies the demand side of the signal (premium and
threshold) while the self-selection half lives in Section 7.3's
observational implications. A D-efficient design with 12 choice tasks of
three alternatives per respondent, and a sample-size target of
\(n = 450\). The target is simulated, not asserted: Appendix C reports a
Monte-Carlo power analysis---simulated respondents with
literature-anchored part-worths and 50\% preference heterogeneity, a
deliberately conservative random design, pooled conditional-logit
estimation, Wald test of the H3 contrast at \(\alpha = .05\)---whose
central results are these. The main effects H1 and H2 are detected with
power at or near 1 from \(n = 250\) upward; the binding constraint is
the interaction. At the anchored medium interaction (half the warranty
main effect, \(\approx\)€75 per day in willingness-to-pay terms), power
is 62\% at \(n = 250\), 72\% at \(n = 300\), and 90\% at \(n = 450\),
with the conventional 80\% threshold crossed near \(n \approx 370\). A
large interaction (\(\approx\)€113/day) is detected with 91\% power
already at \(n = 250\); a small one (\(\approx\)€38/day) is out of reach
at any feasible size---57\% power even at \(n = 900\)---so we state the
design's limit instead of claiming otherwise: at \(n = 450\), the
smallest interaction detectable with 80\% power is about €68 per day.
Twelve tasks rather than ten buy six to nine percentage points of power
at the medium effect and cost no fieldwork. Subgroup analyses by
industry and firm size remain exploratory at this size. Recruitment
through commercial B2B panels operating in the DACH region, with
preregistration (OSF or AsPredicted) \textbf{before} field work,
including the model specification (mixed logit with WTP-space
parameterization; the power analysis uses fixed-parameter conditional
logit with model-based standard errors as its test statistic and prices
the ordinal levels only---the quantified-\(D\) contrasts, the
actuarial-excess estimand, and the kink test are to be added and the
analysis re-run, with respondent-clustered inference, against the final
D-efficient design before registration), exclusion rules, and all
hypotheses below. Estimated cost at €25--60 per completed qualified
interview: roughly €12,000--27,000; timeline including pretest, six to
nine months.

\hypertarget{hypotheses}{%
\subsection{Hypotheses}\label{hypotheses}}

\textbf{H1 (liability premium).} Willingness to pay is increasing in the
warranty attribute---a level effect that insurance value alone predicts
under any risk attitude, and therefore a manipulation check rather than
a test of this theory. The theory-bearing forms are \textbf{H1a}: WTP
exceeds the actuarial value \((1-\hat{q})\varphi D\) by an economically
material margin---the signaling premium is the \emph{excess} over risk
transfer; and \textbf{H1b}: the premium exhibits a kink at
\(\varphi D = v\), the model's threshold.

\textbf{H2 (disclosure discount).} Disclosed AI production lowers WTP
relative to no statement (replication of the disclosure literature in a
B2B expert-services context).

\textbf{H3 (the interaction that matters).} The liability premium is
\emph{larger} under disclosed AI production than under expert
attribution---liability substitutes for the lost origin signal.
Proposition 4 supplies the mechanism conditional on one premise, stated
here as an explicit behavioral assumption rather than a model
implication: AI disclosure lowers the buyer's competence posterior on
the specific offer,
\(\mu_{\mathrm{AI}} < \mu_{\mathrm{expert}}\)---empirically motivated by
the disclosure literature and probed by H2. Under it, the identification
value of the surviving signal is larger under disclosure by exactly
\((\mu_{\mathrm{expert}} - \mu_{\mathrm{AI}})\,\Delta q\, v\). A pure
sentiment account predicts H2 but is silent on H3; the signaling account
predicts H1 unconditionally and H3 exactly insofar as disclosure moves
the competence posterior. H3 is therefore the discriminating hypothesis,
and the experiment is informative \emph{whatever} its sign.

\textbf{H4 (provenance placebo, optional arm).} Adding a ``certified
human-made (content credentials)'' attribute should command little
premium once liability terms are present (Proposition 5); a large
independent provenance premium would count \emph{against} the model's
ranking of signals.

\hypertarget{observational-implications-in-market-data}{%
\subsection{Observational implications in market
data}\label{observational-implications-in-market-data}}

A careful reader proposed that market data could substitute for the
experiment. We take the two proposed tests seriously as
\emph{additional} observable implications, and state why they cannot
carry the identification alone. First, \textbf{premium dynamics}:
Proposition 7 and Figure 6 imply that in AI-exposed, small-ticket
advisory segments, insurers on pooled books should experience
deteriorating loss ratios as AI-polished entrants flow in, followed by
premium increases, tightened underwriting, or forced risk
differentiation; segments with established experience rating should show
widening premium dispersion instead of rising averages. This is a real
prediction, but its identification is confounded by the general
hardening cycle of professional-liability lines, claims-cost inflation,
and regulatory shifts; testing it requires within-insurer, cross-segment
difference-in-differences on data that are largely proprietary. Second,
\textbf{retention--price correlation}: the retained-risk principle
predicts a positive association between chosen deductibles (and rated
premiums) and realized day rates across providers. But deductible choice
is endogenous to wealth, firm size, and risk aversion, all of which also
move prices; without an instrument, the correlation corroborates rather
than identifies. Both tests are worth running; neither replaces the
randomized attribute variation of the conjoint design, which remains the
primary causal test. They complement it on exactly the margin stated
preferences cannot reach: realized, incentivized behavior.

\hypertarget{what-would-falsify-the-theory}{%
\subsection{What would falsify the
theory}\label{what-would-falsify-the-theory}}

We commit to the following: if, in a well-powered preregistered test,
the willingness to pay for warranted expert services does \textbf{not}
exceed the actuarial value of the promised transfer (H1a rejected with a
tight zero on the excess), with no kink at \(\varphi D = v\) (H1b),
while prices for otherwise identical offers vary freely along other
attributes, the central claim of this paper---liability as the surviving
carrier of competence information---is wrong as an economics of this
market, whatever its validity in the laboratory settings of Dulleck et
al.~(2011). Similarly, over the medium run: if realized market prices
for warranted expert services show no premium over unwarranted ones
across several years of AI diffusion while information-only services
hold their prices, the scissors is refuted as a description of the
world. We regard stating this as part of the paper's argument: a theory
about costly commitments should be willing to make one. The
institutional extension adds two sharper edges: if warranted premiums
are \emph{equally} present below and above plausible litigation
thresholds (contradicting Proposition 6), or if providers on pooled
low-deductible insurance sustain the same warranty premium as providers
bearing genuine retained risk (contradicting Proposition 7), the
endogenized model fails even where the baseline might survive.

\hypertarget{conclusion}{%
\section{Conclusion}\label{conclusion}}

A signal is information only insofar as it is expensive to fake.
Generative AI has made the traditional artifacts of expertise cheap to
fake, and the marketplace consequences---compressed premiums for the
previously distinguished, trust discounts on disclosed machine
involvement, rising rhetorical emphasis on ``the human''---are the two
visible blades of a single underlying movement. We have formalized that
movement in a standard signaling model of credence goods, identified the
class of signals that survives it (those whose cost is a contingent
claim on outcomes rather than an input to production), derived the exact
condition under which a warranty restores the separating equilibrium
that artifacts can no longer support (\(\varphi D \geq v\)), shown why
provenance certification cannot do the same, and demonstrated the
dynamics in simulation. None of the mechanism is new; that is its
strength. What is new is the boundary result---\emph{which} signals AI
kills and which it spares, and why the sparing is exact---together with
its institutional fine print: the promise must clear the courthouse's
fixed costs to mean anything (Proposition 6), and it must stay on the
promisor's own balance sheet, or be priced to his own record, to signal
anything (Proposition 7). The empirical intersection this makes urgent
is specified as a preregistered test, flanked by two observational
implications in insurance-market data, with the conditions under which
we would accept defeat.

The practical sentence the theory earns is short. In markets where no
one can tell anymore who did the work, the provider who says \emph{``if
it doesn't solve your problem, I pay''} is not offering a perk. Under
the conditions derived here, that provider is offering the only sentence
the market still believes.

\hypertarget{appendix-a-proofs}{%
\section*{Appendix A: Proofs}\label{appendix-a-proofs}}
\addcontentsline{toc}{section}{Appendix A: Proofs}

\textbf{Proof of Proposition 1.} \emph{(\(\Leftarrow\))} Suppose
\(\eta \geq \eta^{*}\). Consider the strategy profile: \(H\) chooses
\(s_H = a + \Delta q v / k_L \leq \bar{s}\) (feasible by hypothesis),
\(L\) chooses \(s_L = a\) (cost 0); beliefs assign type \(H\) to
\(s \geq s_H\) and \(L\) otherwise; prices \(p(s \geq s_H) = q_H v\),
else \(q_L v\). \(L\)'s best deviation to \(s_H\) yields
\(q_H v - k_L(s_H - a) = q_H v - \Delta q v = q_L v\), no gain;
deviations to \(s \in (a, s_H)\) incur cost without belief change.
\(H\)'s cost at \(s_H\) is \(k_H \Delta q v / k_L < \Delta q v\), so
\(H\) strictly prefers \(s_H\)
(\(q_H v - (k_H/k_L)\Delta q v > q_L v\)). Off-path beliefs are those
specified; participation holds for a positive mass of high types exactly
when \(F(q_H v - (k_H/k_L)\Delta q v) > 0\), which is the statement's
participation clause---absent it, every high type's outside option
exceeds the Riley payoff and no active separating PBE exists even though
the signal is technologically feasible. The profile is then a PBE, and
it is the least-cost separating outcome. \emph{(\(\Rightarrow\))}
Suppose a separating PBE exists with \(H\) at \(s_H\), \(L\) at
\(s_L \neq s_H\), prices \(q_H v\), \(q_L v\) respectively. \(L\)'s
incentive constraint requires
\(q_H v - k_L \max\{0, s_H - a\} \leq q_L v - c_L(s_L) \leq q_L v\) (its
truthful payoff nets out its own artifact cost \(c_L(s_L) \geq 0\),
which only tightens the bound), i.e., \(k_L (s_H - a) \geq \Delta q v\),
i.e., \(s_H \geq a + \Delta q v / k_L\). Feasibility
\(s_H \leq \bar{s}\) then forces
\(\eta = \bar{s} - a \geq \Delta q v / k_L = \eta^{*}\). Hence for
\(\eta < \eta^{*}\) no separating PBE exists, proving (i). For (ii): let
\(s_1, s_2\) be on-path signals carrying posteriors \(\mu_1 > \mu_2\)
and competitive prices \(q_L v + \mu_i \Delta q v\). Since
\(\mu_2 < 1\), Bayes' rule requires \(s_2\) to lie in the support of
\(L\)'s strategy. \(L\)'s optimality at \(s_2\) against a deviation to
\(s_1\) requires
\((\mu_1 - \mu_2) \Delta q v \leq k_L \max\{0, s_1 - a\} - k_L \max\{0, s_2 - a\} \leq k_L \eta\),
hence \(\mu_1 - \mu_2 \leq k_L \eta / (\Delta q v) = \eta / \eta^{*}\).
The bound is a statement about the artifact game: with further
instruments present (Section 3.3), an artifact level can correlate
costlessly with an informative instrument and carry its information as a
label, which is why the statement conditions on prices depending on
\(s\) alone; for continuously mixed strategies the bound is an
essential-supremum statement over on-path signals. (For general strictly
increasing \(c_\theta\) with \(c_\theta(0) = 0\) and \(c_L > c_H\)
pointwise, replace \(k_L(s_H - a)\) by \(c_L(s_H - a)\) throughout: the
threshold becomes \(\eta^{*} = c_L^{-1}(\Delta q\, v)\), and the spread
bound becomes \(c_L(\eta)/(\Delta q\, v)\).) \(\blacksquare\)

\textbf{Proof of Proposition 2.} Under pooling, competitive pricing
gives \(\bar{p}(\lambda) = q_L v + \lambda \Delta q v\) where
\(\lambda\) is the share of \(H\) among active sellers. An \(H\) with
outside option \(r\) participates iff \(\bar{p}(\lambda) \geq r\); all
\(L\) participate. With mass \(\mu_0 F(\bar{p}(\lambda))\) of active
\(H\) and \(1 - \mu_0\) of active \(L\), consistency requires the
fixed-point equation in the text. The map
\(\lambda \mapsto \mu_0 F(\bar{p}(\lambda)) / (\mu_0 F(\bar{p}(\lambda)) + 1 - \mu_0)\)
is nondecreasing from \([0,1]\) to \([0,1]\) (as \(F\) and \(\bar{p}\)
are nondecreasing), so a fixed point exists by Tarski's
theorem---continuity, and hence Brouwer, would require \(F\) continuous,
which we do not assume. Under the restriction
\(\underline{r} > q_L v = \bar{p}(0)\) (invoked for this clause),
\(F(\bar{p}(0)) = 0\) and \(\lambda = 0\) is a fixed point; the
statements concern every fixed point, and the dynamics of Section 5
track the largest. At any fixed point, \(H\)'s payoff is
\(\bar{p}(\lambda) = q_L v + \lambda \Delta q v\), while the separating
payoff is
\(q_H v - (k_H/k_L)\Delta q v = q_L v + (1 - k_H/k_L)\Delta q v\); the
former is strictly smaller iff \(\lambda < 1 - k_H / k_L\).
Participation
\(F(\bar{p}(\lambda)) \leq F(q_H v - (k_H/k_L)\Delta q v)\) follows by
monotonicity of \(F\). For the unraveling clause: the map sends
\([0,1]\) into \([0, \mu_0]\) (set \(F \leq 1\)), so every fixed point
has \(\bar{p}(\lambda) \leq q_L v + \mu_0 \Delta q v\); if \(F\)
vanishes at that price, monotonicity gives \(F(\bar{p}(\lambda)) = 0\),
hence \(T(\lambda) = 0\), and \(\lambda = 0\) is the unique fixed point:
only \(L\) trade, at \(q_L v\). The condition is tight---with a unit
atom at the pooling participation value
(\(\underline{r} = q_L v + \mu_0 \Delta q v\) and
\(F(\underline{r}) = 1\)), both \(\lambda = 0\) and \(\lambda = \mu_0\)
are fixed points, every high type weakly participating in the second; a
smaller atom need not generate a positive fixed point, since
\(T(\mu_0) = \mu_0\) forces \(F(\underline{r}) = 1\). \(\blacksquare\)

\textbf{Proof of Proposition 3.} Assumptions (A1)--(A4) are in force
throughout. Complete the profile of the statement: both active types
choose the free artifact level \(s = a\); \(H\) chooses \(w = 1\), \(L\)
chooses \(w = 0\); \(D\) is a fixed, collectible contract parameter, not
a seller choice. Buyer beliefs, for every artifact level \(s\):
\(\Pr(H \mid s, w = 1) = 1\) and \(\Pr(H \mid s, w = 0) = 0\)---beliefs
condition on the warranty alone. Competitive prices: a warranted offer
believed \(H\) is worth \(q_H v + (1 - q_H)\varphi D\) to the buyer
(value plus expected recovered damages), an unwarranted offer believed
\(L\) is worth \(q_L v\). \(H\)'s payoff:
\(p_H - (1 - q_H)\varphi D = q_H v\). \(L\)'s payoff from honesty:
\(q_L v\). \(L\)'s deviation to \(w = 1\) nets, relative to honesty,
\(p_H - (1 - q_L)\varphi D - q_L v = \Delta q\,(v - \varphi D)\):
strictly negative for \(\varphi D > v\), zero at the boundary, positive
below---deterrence is exactly \(\varphi D \geq v\), strict off the
boundary, and it holds by hypothesis. \(H\)'s deviation to \(w = 0\)
yields \(q_L v < q_H v\), unprofitable. Artifact deviations leave
beliefs unchanged by construction and only add cost, so none is
profitable. Existence holds for every \(\eta \geq 0\) since no step used
\(a\), \(k_\theta\), or \(\bar{s}\); the cost differential is
\(\kappa_L - \kappa_H = \Delta q \varphi D\), independent of \(a\).
(Selection: where both artifact- and liability-separation exist, \(H\)
weakly prefers liability, whose net signaling cost is zero versus
\((k_H / k_L)\Delta q v > 0\); D1 eliminates pooling when a separating
deviation with these payoffs exists.) \(\blacksquare\)

\textbf{Proof of Proposition 4.}
\(\Pi(\lambda) = q_H v - \bar{p}(\lambda) = (1 - \lambda)\Delta q v\);
\(\Pi'(\lambda) = -\Delta q v < 0\). Entry of mass \(e \geq 0\) of
low-type supply shifts the participation map
\(T(\lambda; e) = \mu_0(e) F(\bar{p}(\lambda)) / (\mu_0(e) F(\bar{p}(\lambda)) + 1 - \mu_0(e))\),
which is nondecreasing in \(\lambda\) and nonincreasing in \(e\) (entry
lowers \(\mu_0(e)\)); by monotone comparative statics on the complete
lattice (Topkis), the largest fixed point \(\lambda^{*}(e)\) is
nonincreasing in \(e\), so \(\Pi\) is nondecreasing in \(e\). If
\(A \mapsto a(A)\) is nondecreasing and \(A \mapsto e(A)\)
nondecreasing, then (i) follows from Proposition 1 once
\(\eta(A) = \bar{s} - a(A) < \eta^{*}\), and (ii) from the composition.
\(\blacksquare\)

\textbf{Proof of Proposition 5.} Let signal \(\varsigma \in \{0,1\}\)
cost \(m \geq 0\) for both types. Type \(\theta\)'s payoff difference
between \(\varsigma = 1\) and \(\varsigma = 0\) is \(p(1) - p(0) - m\)
for \emph{both} types (origin does not enter \(q_\theta\) by assumption,
so no other payoff term differs). (i) Pure separation with \(H\)
certified and \(L\) not requires both incentive constraints, which pin
\(p(1) - p(0) = \Delta q v = m\) and give both types the payoff
\(q_L v\). The knife edge is closed by participation, not by strictness:
every high type's outside option exceeds \(q_L v\) (by the hypothesis of
part (i)), so no high type is active in such a profile and no separating
PBE with active high types exists; away from \(m = \Delta q v\) the
common deviation calculus makes both types strictly prefer the same
action; the reverse assignment---\(L\) certified, \(H\) not---fails
immediately, since \(L\)'s incentive constraint would require
\(-m \geq \Delta q\, v\). (ii) Informativeness with both certificate
outcomes on path forces indifference of both types---they share one
deviation calculus---hence \(p(1) - p(0) = m\) and a posterior gap of
exactly \(t = m/(\Delta q\, v)\). Writing \(\omega\) for the aggregate
certification probability, Bayes plausibility pins the posteriors to
\(\mu(1) = \mu + (1-\omega)t\) and \(\mu(0) = \mu - \omega t\), and the
type strategies to \(\alpha_H = \omega\mu(1)/\mu\),
\(\alpha_L = \omega(1-\mu(1))/(1-\mu)\), admissible iff
\(0 \leq \mu - \omega t\) and \(\mu + (1-\omega)t \leq 1\). Conversely,
every admissible pair \((\mu, \omega)\) defines such a PBE with common
payoff \(u(\mu, \omega) = p(0) = q_L v + \mu\,\Delta q\, v - \omega m\);
the slice \(\omega = \mu\) admits a closed form but is not exhaustive:
equilibria with small \(\omega\) burn little and can clear outside
options the slice cannot. With participation endogenous, high types with
\(r \leq u\) enter, so \((\mu, \omega)\) must also solve the fixed point
in the statement; and since
\(u(\mu, \omega) < q_L v + \mu_0 \Delta q\, v\) strictly for every
\(\omega > 0\) (as \(\mu \leq \mu_0\)), an \(\underline{r}\) weakly
above that cap leaves no high type willing to enter in any informative
profile: no certificate is informative for any \(m\)---the same
outside-option floor that kills pure separation. (iii) In every
informative equilibrium of (ii) each type's payoff is
\(u(\mu, \omega) < q_L v + \mu\,\Delta q\, v\) for \(\omega > 0\). The
dominating equilibrium is a no-certificate pooling equilibrium of the
same game---not a profile with an artificially frozen population, which
with endogenous entry need not be an equilibrium at all: dropping the
certificate raises payoffs and can pull further high types in. Instead,
let
\(T_0(\lambda) = \mu_0 F(\bar{p}(\lambda))/(\mu_0 F(\bar{p}(\lambda)) + 1 - \mu_0)\)
be the no-certificate participation map:
\(\bar{p}(\mu) > u(\mu, \omega)\) and monotonicity give
\(T_0(\mu) \geq \mu\), so \(T_0\) maps \([\mu, \mu_0]\) into itself and
has a fixed point \(\lambda^{*} \geq \mu\) (Tarski). In that pooling
equilibrium every seller active in the certificate equilibrium earns
\(q_L v + \lambda^{*} \Delta q\, v \geq q_L v + \mu \Delta q\, v > u(\mu, \omega)\)---strictly
more. We adopt the selection convention of discarding equilibria
strictly Pareto-dominated for every sender type by another equilibrium
of the same game---an additional modeling assumption, implied neither by
PBE nor by D1, undefeatedness (Mailath, Okuno-Fujiwara, \& Postlewaite,
1993), or the persuasiveness criterion (Zeng, 2025), whose verification
we do not undertake here. Under it, the certificate is off path or
uninformative in every retained equilibrium, and on-path posterior
beliefs equal the prior conditional on participation. \(\blacksquare\)

\textbf{Proof of Proposition 6.} Define
\(g(v) \equiv \rho\,\delta\, v - (1-\rho)\,C_L(\delta v) = \rho\delta v - (1-\rho)\left(c_0 + c_1 (\delta v)^{\alpha}\right)\)
on \([0, \infty)\). Then \(g(0) = -(1-\rho)c_0 < 0\);
\(g''(v) = -(1-\rho)\,c_1\,\alpha(\alpha - 1)\,\delta^{\alpha} v^{\alpha - 2} > 0\)
for \(\alpha \in (0,1)\), so \(g\) is strictly convex; and
\(g(v) \to \infty\) as \(v \to \infty\) since the linear term dominates
(\(\alpha < 1\)). A strictly convex function that starts negative and
diverges to \(+\infty\) crosses zero exactly once; call the root
\(v_{\min}\), with \(g < 0\) below and \(g > 0\) above. Credibility of
suit at ticket \(v\) is exactly \(g(v) > 0\), which under strict
convexity holds iff \(v > v_{\min}\) and fails at \(v = v_{\min}\)
(where \(g = 0\)) and below---hence the strict inequality in the
statement. Given credibility, settlement in the shadow of judgment
transfers \(\rho D = \rho\delta v\) in expectation upon a failure the
client believes occurred; the mimicry-deterrence algebra of Proposition
3 with \(\varphi D\) replaced by \(\rho\,\delta v\) requires
\(\rho\delta v \geq v\), i.e., \(\rho\delta \geq 1\): strict for
\(\rho\delta > 1\), with equality---and hence weak separation, the mimic
exactly indifferent---at \(\rho\delta = 1\) wherever \(g(v) > 0\); the
condition is independent of \(v\). Hence for \(\rho\delta > 1\)
separation is feasible iff \(v > v_{\min}\); for \(\rho\delta < 1\),
deterrence fails at every \(v\) regardless of credibility. The strict
convexity of \(g\) uses \(c_1 > 0\), now explicit in the setup.
Comparative statics by the implicit function theorem:
\(\partial v_{\min}/\partial c_0 = (1-\rho)/g'(v_{\min}) > 0\) and
\(\partial v_{\min}/\partial \rho = -\left(\delta v_{\min} + C_L(\delta v_{\min})\right)/g'(v_{\min}) < 0\),
using \(g'(v_{\min}) > 0\) (the function crosses from below at its
unique root, and strict convexity with \(g(0)<0\) forces a positive
slope at the crossing). \(\blacksquare\)

\textbf{Proof of Proposition 7.} Beliefs: warranty \(\Rightarrow H\).
Competitive price of a warranted offer believed \(H\):
\(p_H = q_H v + (1 - q_H)\varphi D\) (the buyer collects \(D\) upon
verified failure regardless of who ultimately funds it). (i)
\emph{Pooled premiums.} In the candidate separating equilibrium only
high types insure, so the competitive zero-loading premium on the
insured book is \(\pi = (1 - q_H)\varphi(D - SB)\), charged identically
to any buyer of cover. A type-\(\theta\) warranted seller's total
outcome-contingent cost is then \(\pi + (1 - q_\theta)\varphi SB\), and
\(H\)'s net payoff is \(p_H - \pi - (1 - q_H)\varphi SB = q_H v\), so
high-type participation holds as in Proposition 3. \(L\)'s deviation to
the warranty yields
\(p_H - \pi - (1 - q_L)\varphi SB = q_H v + (1 - q_H)\varphi SB - (1 - q_L)\varphi SB = q_H v - \Delta q\,\varphi\, SB\)
versus honesty \(q_L v\): deterrence holds iff
\(\Delta q\, \varphi\, SB \geq \Delta q\, v\), i.e.,
\(\varphi SB \geq v\). The promised \(D\) cancels because the insurer
pays \(D - SB\) for both types at the same premium. The pricing tie is
load-bearing, not cosmetic: for an arbitrary common \(\pi\), \(L\)'s
incentive constraint reads
\(\Delta q\, v \leq \pi + (1 - q_L)\varphi SB - (1 - q_H)\varphi D\), so
an overpriced pool can deter the mimic at small \(SB\) (while destroying
\(H\)'s participation) and an underpriced one can subsidize him at large
\(SB\)---which is exactly the subsidized-mimicry configuration the
simulation exhibits in Section 5.6, and the reason the statement fixes
the premium at the competitive level. Throughout, \(0 \leq SB \leq D\)
per the setup; for \(SB > D\) every expression applies to the retained
amount \(\min\{SB, D\}\). All quantities are the collectible ones: with
seller wealth \(S\), the borne retention is \(R = \min\{SB, S\}\) and
the buyer prices the collectible transfer, so replacing \(SB\) by \(R\)
and \(D\) by \(R + C\) leaves the algebra unchanged---and the
substitution has bite: a nominal \(\varphi SB \geq v\) with
\(S + I < v\) leaves mimicry strictly profitable, so no admissible
contract separates, however large the printed deductible. (ii)
\emph{Full experience rating.}
\(\pi_\theta = (1 - q_\theta)\varphi(D - SB)\), so type \(\theta\)'s
total outcome-contingent cost is
\((1 - q_\theta)\varphi(D - SB) + (1 - q_\theta)\varphi SB = (1 - q_\theta)\varphi D\):
the differential is \(\Delta q\,\varphi D\), and Proposition 3's
condition \(\varphi D \geq v\) applies verbatim. Intermediate
credibility weights \(z \in (0,1)\) on own experience yield
differentials \(\Delta q\,\varphi\,(SB + z(D - SB))\), monotone in \(z\)
between the two poles. Judgment-proofness caps: an uninsured seller can
transfer at most \(\min\{D, S\}\), an insured one \(\min\{D, S + I\}\);
all conditions apply to the transferable amount. \(\blacksquare\)

\hypertarget{appendix-b-simulation-details}{%
\section*{Appendix B: Simulation
Details}\label{appendix-b-simulation-details}}
\addcontentsline{toc}{section}{Appendix B: Simulation Details}

The simulation is implemented in Python (NumPy/pandas), fixed seed, and
accompanies this paper. Per period and replication: (1) the AI frontier
\(a_t\) updates on the logistic path; (2) active sellers post signals
per the regime's equilibrium logic (artifact separation while
\(\eta_t \geq \eta^{*}\); otherwise pooling in Regime A / warranty
separation in Regime B); (3) buyers price competitively given beliefs;
(4) outcomes draw binomially, verified failures pay \(D\); (5) low types
update mimicry propensity by realized profit comparison
(\(\varepsilon\)-greedy with decay on losses, initial
\(\varepsilon = 0.05\), decay 0.97 on loss signal, belief smoothing
0.3); (6) high types update smoothed payoff expectations (weight 0.3)
and exit/enter against their outside option. Reported aggregates are
means and 2.5/97.5 percentiles across 500 replications. The phase map
(Figure 3) evaluates, at each \((\eta, \varphi D / v)\) grid point, the
frequency over 40 reduced runs in which separation survives
finite-sample mimicry tests (30 periods, 300 sellers). Figure 4 samples
200 finite markets of 800 sellers per composition point. The litigation
module (Figure 5) draws match-level provability
\(\rho \sim \mathrm{Beta}(24\rho_0, 24(1-\rho_0))\) and evaluates
credibility and deterrence per match over 40 log-spaced ticket sizes
(200 replications × 400 matches; \(c_0 = 2000\), \(c_1 = 7.9\),
\(\alpha = 0.66\), \(\delta = 2\)). The insurance module (Figures 6 and
7(d)) runs 96 months × 200 replications with adaptive low-type warranty
adoption (logistic on smoothed realized profit gap, slope 80, memory
0.9), insurer updating of verified-claim frequencies (EMA 0.85/0.15),
credibility-weighted individual rating (\(z_i = n_i/(n_i + 20)\)), and
buyer belief updating from realized success rates (EMA 0.9/0.1).
Premiums are computed as \((D - SB)\) times the relevant verified-claim
frequency---pooled or credibility-weighted individual---so that in
expectation \(\pi_\theta = (1 - q_\theta)\varphi(D - SB)\), Proposition
7's actuarial benchmark. (An earlier code version scaled premiums by an
additional factor \(\varphi\); the correction strengthens the rated
regime's recovery relative to previously circulated figures.) Full
parameter listing in Table 2; code files:
\texttt{simulation\_advanced.py} (all experiments; contains the baseline
of \texttt{simulation.py} verbatim), with visualization scripts
\texttt{visualize\_baseline.py}, \texttt{visualize\_advanced.py}, and
\texttt{visualize\_combined.py}, the power-analysis script
\texttt{power\_cbc.py} of Appendix C, and \texttt{boundary\_checks.py},
which verifies the knife-edge and counterexample calculations behind
Propositions 2--7 and the friction figures of Sections 3--6
(deterministic, runtime under one second). All code files are archived
at https://doi.org/10.6084/m9.figshare.33106946.

\hypertarget{appendix-c-power-simulation-for-the-proposed-experiment}{%
\section*{Appendix C: Power Simulation for the Proposed
Experiment}\label{appendix-c-power-simulation-for-the-proposed-experiment}}
\addcontentsline{toc}{section}{Appendix C: Power Simulation for the
Proposed Experiment}

The sample-size target of Section 7.1 replaces bare adequacy language
with a simulation. The analysis is implemented in \texttt{power\_cbc.py}
(NumPy, fixed seed), archived with the simulation code of Appendix B.

\textbf{Setup.} Design as in Section 7.1: five attributes---warranty (3
levels), disclosure (3), price (4 levels at €1,000/1,200/1,400/1,600 per
day, bracketing the BDU (2025) anchor), references (2), turnaround
(2)---twelve choice tasks of three alternatives, forced choice. Profiles
are drawn uniformly at random with exact within-task duplicates redrawn:
a \emph{random} design, conservative relative to the D-efficient design
to be fielded. Simulated respondents draw individual part-worths from
normal distributions around literature-anchored means---warranty full
0.75 (\(\approx\)€150/day at a price coefficient of \(-0.005\) per euro,
roughly 11\% of the anchor day rate and the order of magnitude of the
B2B guarantee premia in McColl et al., 2019), warranty mid 0.40, AI
disclosure \(-0.50\), expert attribution +0.25, references 0.35,
turnaround 0.25---with heterogeneity standard deviations of 50\% of each
mean and a lognormal price coefficient (log-sd 0.30, mean-one
multiplier). The H3 interaction---the \emph{additional} warranty-full
premium under disclosed AI production relative to expert
attribution---is varied at 25\%, 50\%, and 75\% of the warranty main
effect (\(\approx\)€38, €75, €113 per day), with the mid-warranty
interaction at half of it. Each replication simulates choices, fits a
pooled conditional logit with the full interaction structure (11
parameters, Newton--Raphson), and evaluates the Wald test of the H3
contrast at \(\alpha = .05\), two-sided with correct sign required; 400
replications per cell. Parameter recovery was verified at
\(n = 4{,}000\): the pooled estimator mildly attenuates utilities under
heterogeneity, as expected, but recovers willingness-to-pay ratios
essentially without bias, so the power figures refer to the quantity of
interest.

\textbf{Results.} Power for H1 (warranty main effect) is 1.00 in every
cell; for H2 (disclosure discount) it is \(\geq\) .87 at \(n = 150\) and
\(\geq\) .98 from \(n = 250\). For the H3 contrast:

\begin{longtable}[]{@{}
  >{\raggedleft\arraybackslash}p{(\columnwidth - 8\tabcolsep) * \real{0.1724}}
  >{\centering\arraybackslash}p{(\columnwidth - 8\tabcolsep) * \real{0.2069}}
  >{\centering\arraybackslash}p{(\columnwidth - 8\tabcolsep) * \real{0.2069}}
  >{\centering\arraybackslash}p{(\columnwidth - 8\tabcolsep) * \real{0.2069}}
  >{\centering\arraybackslash}p{(\columnwidth - 8\tabcolsep) * \real{0.2069}}@{}}
\caption{Simulated power for the H3 contrast (Wald test,
\(\alpha = .05\), correct sign required), twelve tasks, 400 replications
per cell. MDES: smallest interaction detectable with 80\% power, by
linear interpolation.}\tabularnewline
\toprule\noalign{}
\begin{minipage}[b]{\linewidth}\raggedleft
\(n\)
\end{minipage} & \begin{minipage}[b]{\linewidth}\centering
H3 \(\approx\) €38/day
\end{minipage} & \begin{minipage}[b]{\linewidth}\centering
H3 \(\approx\) €75/day
\end{minipage} & \begin{minipage}[b]{\linewidth}\centering
H3 \(\approx\) €113/day
\end{minipage} & \begin{minipage}[b]{\linewidth}\centering
MDES (80\% power)
\end{minipage} \\
\midrule\noalign{}
\endfirsthead
\toprule\noalign{}
\begin{minipage}[b]{\linewidth}\raggedleft
\(n\)
\end{minipage} & \begin{minipage}[b]{\linewidth}\centering
H3 \(\approx\) €38/day
\end{minipage} & \begin{minipage}[b]{\linewidth}\centering
H3 \(\approx\) €75/day
\end{minipage} & \begin{minipage}[b]{\linewidth}\centering
H3 \(\approx\) €113/day
\end{minipage} & \begin{minipage}[b]{\linewidth}\centering
MDES (80\% power)
\end{minipage} \\
\midrule\noalign{}
\endhead
\bottomrule\noalign{}
\endlastfoot
150 & .12 & .43 & .75 & \textgreater{} €113 \\
250 & .17 & .62 & .91 & \(\approx\) €98 \\
300 & .21 & .72 & .94 & \(\approx\) €89 \\
450 & .34 & .90 & 1.00 & \(\approx\) €68 \\
600 & .43 & .93 & 1.00 & \(\approx\) €65 \\
900 & .57 & .99 & 1.00 & \(\approx\) €58 \\
\end{longtable}

Ten tasks instead of twelve cost six to nine percentage points of power
at the medium effect (.56 vs.~.62 at \(n = 250\); .63 vs.~.72 at
\(n = 300\)), which is why twelve are specified. Three limits deserve
statement. First, conditional logit is the test statistic---standard in
CBC power practice and fast enough for dense replication---but its Wald
test rests on the pooled information matrix, which ignores
within-respondent correlation across the twelve tasks and therefore, if
anything, overstates power at the interaction; the preregistered
estimator (mixed logit in WTP space) models the heterogeneity the
simulation injects, preregistered inference will use
respondent-clustered standard errors, and the analysis is to be re-run
against the final D-efficient design under that variance estimator (the
design change can only improve on the random design used here; the
variance correction works the other way, which is one more reason the
target is \(n = 450\) rather than the smallest passing cell). Second,
the assumed effect sizes are anchors, not measurements: the simulation
converts the sample-size claim into a conditional statement---\emph{if}
effects are of the anchored magnitudes, \emph{then} \(n = 450\) detects
the discriminating interaction with 90\% power---whose antecedent is
exactly what the experiment tests. Third, an interaction below roughly
€70/day would likely go undetected at this size; a null on H3 is
therefore informative against \emph{large} substitution effects only,
and the preregistration will state this bound explicitly.

\hypertarget{use-of-generative-ai}{%
\section*{Use of Generative AI}\label{use-of-generative-ai}}
\addcontentsline{toc}{section}{Use of Generative AI}

In preparing this manuscript the author used Anthropic's Claude,
including Claude Code (accessed July 2026), for four purposes:
supplementary literature search and source discovery; drafting and
language editing of the text; implementing and running the numerical and
simulation code underlying the model calculations and the figures; and
assistance with the derivations. The research question, the model and
its assumptions, the interpretation of the results and all final
decisions about content are the author's. AI technology was not used to
generate, alter or fabricate research data, and is not credited as an
author. No AI-assisted output entered the manuscript unverified: every
reference was checked against the publisher record, and every reported
number is regenerated from the deposited replication code rather than
transcribed (see Data availability). The author takes full
responsibility for the content of the manuscript, including all claims,
citations and code.

\hypertarget{data-availability}{%
\section*{Data availability}\label{data-availability}}
\addcontentsline{toc}{section}{Data availability}

Simulation code, parameters, random seeds and all numerical results
underlying every figure and every reported number are openly available
at https://doi.org/10.6084/m9.figshare.33106946.

\hypertarget{references}{%
\section*{References}\label{references}}
\addcontentsline{toc}{section}{References}

\leavevmode\vadjust pre{\hypertarget{refs}{}}%
\begin{CSLReferences}{0}{0}
Akerlof, G. A. (1970). The market for ``lemons'': Quality uncertainty
and the market mechanism. \emph{Quarterly Journal of Economics, 84}(3),
488--500. https://doi.org/10.2307/1879431

AVB-V (current wordings, 2009--2025). \emph{Allgemeine
Versicherungsbedingungen zur Haftpflichtversicherung für
Vermögensschäden.} Standard German market wordings descending from the
non-binding model conditions of the German Insurance Association (GDV);
§ 4 no. 2 excludes claims ``soweit sie auf Grund Vertrages oder
besonderer Zusage über den Umfang der gesetzlichen Haftpflicht
hinausgehen.'' A materially identical clause appears in Austrian
standard wordings. {[}Primary market documents; clause wording verified
against published insurer conditions dated 2009, 2020, and May 2025;
checked August 2026.{]}

Banks, J. S., \& Sobel, J. (1987). Equilibrium selection in signaling
games. \emph{Econometrica, 55}(3), 647--661.

Bauer, A. (2026a). \emph{The fallback as signal: Preserved human skill,
liability, and competence signaling in credence-good markets under
improving AI.} arXiv preprint arXiv:2608.04276. {[}Preprint, not
peer-reviewed; companion paper endogenizing the competence margin of
Section 6.2.{]}

Bauer, A. (2026b). \emph{The institutional window: Occupation- and
jurisdiction-specific calibration of liability signaling for preserved
human fallback capability.} arXiv preprint arXiv:2608.05969.
{[}Preprint, not peer-reviewed; companion paper building on the present
framework, calibrating the enforceability map of Section 4.1.{]}

BDU --- Bundesverband Deutscher Unternehmensberatungen (2025).
\emph{Honorare im Consulting 2025}. Bonn: BDU. {[}Industry study; not
peer-reviewed.{]}

Boulding, W., \& Kirmani, A. (1993). A consumer-side experimental
examination of signaling theory: Do consumers perceive warranties as
signals of quality? \emph{Journal of Consumer Research, 20}(1),
111--123. https://doi.org/10.1086/209337

Brynjolfsson, E., Chandar, B., \& Chen, R. (2025). \emph{Canaries in the
coal mine? Six facts about the recent employment effects of artificial
intelligence.} Working paper, version November 2025. {[}Not
peer-reviewed; figures preliminary.{]}

Buder, F., \& Unfried, M. (2024). Transparency without trust: The impact
of consumer skepticism of AI-generated marketing content. \emph{NIM
INSIGHTS Research Magazine, 7.} Nürnberg Institut für
Marktentscheidungen. {[}Institute research publication, not
peer-reviewed; survey of n = 1,000 per country (USA/UK/Germany); the
labeling experiment itself n = 298 (Germany); consumer context.{]}

C2PA --- Coalition for Content Provenance and Authenticity (2026).
\emph{Content Credentials: C2PA Technical Specification, Version 2.4.}
spec.c2pa.org. {[}Current published version, verified August 2026;
predecessor v2.3 carries a changelog date of December 2025 and a
document date of January 5, 2026. The specification's own explainer
distinguishes provenance and authenticity from claims of truth or
competence; Proposition 5's assumption---type-independent cost, no
effect on \(q_\theta\)---is version-independent.{]}

Cho, I.-K., \& Kreps, D. M. (1987). Signaling games and stable
equilibria. \emph{Quarterly Journal of Economics, 102}(2), 179--221.

Cho, I.-K., \& Sobel, J. (1990). Strategic stability and uniqueness in
signaling games. \emph{Journal of Economic Theory, 50}(2), 381--413.

Connelly, B. L., Certo, S. T., Ireland, R. D., \& Reutzel, C. R. (2011).
Signaling theory: A review and assessment. \emph{Journal of Management,
37}(1), 39--67. https://doi.org/10.1177/0149206310388419

Connelly, B. L., Certo, S. T., Reutzel, C. R., DesJardine, M. R., \&
Zhou, Y. S. (2025). Signaling theory: State of the theory and its
future. \emph{Journal of Management, 51}(1), 24--61.
https://doi.org/10.1177/01492063241268459

Cooper, R., \& Ross, T. W. (1985). Product warranties and double moral
hazard. \emph{RAND Journal of Economics, 16}(1), 103--113.

Cooter, R. D., \& Rubinfeld, D. L. (1989). Economic analysis of legal
disputes and their resolution. \emph{Journal of Economic Literature,
27}(3), 1067--1097.

Darby, M. R., \& Karni, E. (1973). Free competition and the optimal
amount of fraud. \emph{Journal of Law and Economics, 16}(1), 67--88.
https://doi.org/10.1086/466756

Dulleck, U., \& Kerschbamer, R. (2006). On doctors, mechanics, and
computer specialists: The economics of credence goods. \emph{Journal of
Economic Literature, 44}(1), 5--42.
https://doi.org/10.1257/002205106776162717

Dulleck, U., Kerschbamer, R., \& Sutter, M. (2011). The economics of
credence goods: An experiment on the role of liability, verifiability,
reputation, and competition. \emph{American Economic Review, 101}(2),
526--555. https://doi.org/10.1257/aer.101.2.526

Erlei, A. (2025). \emph{From digital distrust to codified honesty:
Experimental evidence on generative AI in credence goods markets.} arXiv
preprint arXiv:2509.06069. {[}Preprint, not peer-reviewed; one-shot
experiments with human and LLM participants.{]}

Erlei, A., \& Meub, L. (2026). \emph{LLM-agent interactions on markets
with information asymmetries.} arXiv preprint arXiv:2603.08853.
{[}Preprint, not peer-reviewed; simulation study with LLM agents, no
human subjects.{]}

European Union (2024/2026). Regulation (EU) 2024/1689 (Artificial
Intelligence Act), Article 50, as amended by Regulation (EU) 2026/1744
(in force July 27, 2026; Article 50(2) transition for pre-existing
generative systems until December 2, 2026). \emph{Official Journal of
the European Union.} {[}Deadlines verified against the amending
regulation, August 2026.{]}

Hui, X., Reshef, O., \& Zhou, L. (2024). The short-term effects of
generative artificial intelligence on employment: Evidence from an
online labor market. \emph{Organization Science, 35}(6).
https://doi.org/10.1287/orsc.2023.18441

Humlum, A., \& Vestergaard, E. (2025). \emph{Large language models,
small labor market effects.} NBER Working Paper No.~33777.
https://www.nber.org/papers/w33777 {[}Not peer-reviewed.{]}

Kirmani, A., \& Rao, A. R. (2000). No pain, no gain: A critical review
of the literature on signaling unobservable product quality.
\emph{Journal of Marketing, 64}(2), 66--79.
https://doi.org/10.1509/jmkg.64.2.66.18000

Köbis, N., Rahwan, Z., Rilla, R., Supriyatno, B. I., Bersch, C., Ajaj,
T., Bonnefon, J.-F., \& Rahwan, I. (2025). Delegation to artificial
intelligence can increase dishonest behaviour. \emph{Nature, 646}(8083),
126--134. https://doi.org/10.1038/s41586-025-09505-x

Mailath, G. J. (1987). Incentive compatibility in signaling games with a
continuum of types. \emph{Econometrica, 55}(6), 1349--1365.

Mailath, G. J., Okuno-Fujiwara, M., \& Postlewaite, A. (1993).
Belief-based refinements in signalling games. \emph{Journal of Economic
Theory, 60}(2), 241--276. https://doi.org/10.1006/jeth.1993.1043

McColl, R., Truong, Y., \& La Rocca, A. (2019). Service guarantees as a
base for positioning in B2B. \emph{Industrial Marketing Management, 81},
78--86. https://doi.org/10.1016/j.indmarman.2018.11.015 {[}Volume and
pages verified via the corresponding author's institutional CV; the
``81(2)'' circulating on aggregators is an artifact.{]}

OGH --- Oberster Gerichtshof (Austria). Rechtssatz RS0081898 (line of
authority since 7 Ob 145/66 of September 28, 1966; register entries
through October 2025): liability insurance does not cover the insured's
contractual performance interest. Rechtsinformationssystem des Bundes
(RIS). {[}Case-law register; checked August 2026.{]}

PwC --- PricewaterhouseCoopers (2026). \emph{Global AI Jobs Barometer
2026.} Released June 15, 2026; analysis of more than one billion job
advertisements across 27 countries and territories.
https://www.pwc.com/gx/en/news-room/press-releases/2026/pwc-2026-ai-jobs-barometer.html
{[}Industry study, not peer-reviewed; job-advertisement data.{]}

Ramey, G. (1996). D1 signaling equilibria with multiple signals and a
continuum of types. \emph{Journal of Economic Theory, 69}(2), 508--531.

Rix, J., Berger, B., Hess, T., \& Rzepka, C. (2025). The algorithm
discount: Explaining consumers' valuation of human- versus
algorithm-created digital products. \emph{Journal of Management
Information Systems, 42}. https://doi.org/10.1080/07421222.2025.2487308

Rothschild, M., \& Stiglitz, J. (1976). Equilibrium in competitive
insurance markets: An essay on the economics of imperfect information.
\emph{Quarterly Journal of Economics, 90}(4), 629--649.
https://doi.org/10.2307/1885326

Schilke, O., \& Reimann, M. (2025). The transparency dilemma: How AI
disclosure erodes trust. \emph{Organizational Behavior and Human
Decision Processes, 188}, 104405.
https://doi.org/10.1016/j.obhdp.2025.104405

Shavell, S. (1982). Suit, settlement, and trial: A theoretical analysis
under alternative methods for the allocation of legal costs.
\emph{Journal of Legal Studies, 11}(1), 55--81.
https://doi.org/10.1086/467692

Shavell, S. (1986). The judgment proof problem. \emph{International
Review of Law and Economics, 6}(1), 45--58.
https://doi.org/10.1016/0144-8188(86)90038-4

Sikhondze, B. L., Ye, H., Wang, W., Zhan, X., \& Santhanam, R. (2025).
Are you willing to pay for generative artificial intelligence (GenAI)
products? Disentangling the disclosure effects and the mediating role of
psychological value. \emph{Journal of Management Information Systems,
42}(2). https://doi.org/10.1080/07421222.2025.2487314

Spence, M. (1973). Job market signaling. \emph{Quarterly Journal of
Economics, 87}(3), 355--374. https://doi.org/10.2307/1882010

Sultana, N., Islam, M. R., Hossain, M. T., \& Wasi, A. T. (2026).
\emph{Struggle premium: How human effort and imperfection drive
perceived value in the age of AI.} arXiv preprint arXiv:2604.15324.
{[}Preprint, not peer-reviewed; small convenience sample; adjacent
(creative/consumer) market.{]}

von Wedel, P., \& Hagist, C. (2022). Physicians' preferences and
willingness to pay for artificial intelligence-based assistance tools: A
discrete choice experiment among German radiologists. \emph{BMC Health
Services Research, 22}. https://doi.org/10.1186/s12913-022-07769-x

Zeng, H. (2025). \emph{Persuasive selection in signaling games.} arXiv
preprint arXiv:2511.00718. {[}Preprint, not peer-reviewed.{]}

\end{CSLReferences}

\end{document}